\documentclass[twocolumn]{aastex631}

\newcommand{\hydro}{\texttt{HyDRo}\,}

\begin{document}

\title{Rocky planet or water world? Observability of low-density lava world atmospheres}

\correspondingauthor{Anjali A. A. Piette}
\email{apiette@carnegiescience.edu}

\author[0000-0002-4487-5533]{Anjali A. A. Piette}
\affiliation{Earth and Planets Laboratory \\ Carnegie Institution for Science \\ Washington, D.C., 20015, USA}

\author[0000-0002-8518-9601]{Peter Gao}
\affiliation{Earth and Planets Laboratory \\ Carnegie Institution for Science \\ Washington, D.C., 20015, USA}

\author[0000-0003-3913-296X]{Kara Brugman}
\affiliation{Earth and Planets Laboratory \\ Carnegie Institution for Science \\ Washington, D.C., 20015, USA}
\affiliation{Facility for Open Research in a Compressed Environment \\ Arizona State University \\ Tempe, AZ 85287, USA}

\author{Anat Shahar}
\affiliation{Earth and Planets Laboratory \\ Carnegie Institution for Science \\ Washington, D.C., 20015, USA}

\author[0000-0002-3286-7683]{Tim Lichtenberg}
\affiliation{Kapteyn Astronomical Institute, University of Groningen, P.O. Box 800, 9700 AV Groningen, The Netherlands}

\author[0000-0001-8926-4122]{Francesca Miozzi}
\affiliation{Earth and Planets Laboratory \\ Carnegie Institution for Science \\ Washington, D.C., 20015, USA}

\author[0000-0001-6241-3925]{Peter Driscoll}
\affiliation{Earth and Planets Laboratory \\ Carnegie Institution for Science \\ Washington, D.C., 20015, USA}



\begin{abstract}

Super-Earths span a wide range of bulk densities, indicating a diversity in interior conditions beyond that seen in the solar system. In particular, an emerging population of low-density super-Earths may be explained by volatile-rich interiors. Among these, low-density lava worlds have dayside temperatures high enough to evaporate their surfaces, providing a unique opportunity to probe their interior compositions and test for the presence of volatiles. In this work, we investigate the atmospheric observability of low-density lava worlds. We use a radiative-convective model to explore the atmospheric structures and emission spectra of these planets, focusing on three case studies with high observability metrics and sub-stellar temperatures spanning $\sim$1900--2800~K: HD~86226~c, HD~3167~b and 55~Cnc~e. Given the possibility of mixed volatile and silicate interior compositions for these planets, we consider a range of mixed volatile and rock vapor atmospheric compositions. This includes a range of volatile fractions and three volatile compositions: water-rich (100\% H$_2$O), water with CO$_2$ (80\% H$_2$O+20\% CO$_2$), and a desiccated O-rich scenario (67\% O$_2$+33\%CO$_2$). We find that spectral features due to H$_2$O, CO$_2$, SiO and SiO$_2$ are present in the infrared emission spectra as either emission or absorption features, depending on dayside temperature, volatile fraction and volatile composition. We further simulate JWST secondary eclipse observations for each of the three case studies, finding that H$_2$O and/or CO$_2$ could be detected with as few as $\sim$5 eclipses. Detecting volatiles in these atmospheres would provide crucial independent evidence that volatile-rich interiors exist among the super-Earth population.

\end{abstract}

\keywords{Exoplanets -- Exoplanet structure -- Exoplanet atmospheres -- Super-Earths}


\section{Introduction} \label{sec:intro}

\begin{figure*}
    \centering
    \includegraphics[width=0.5\textwidth]{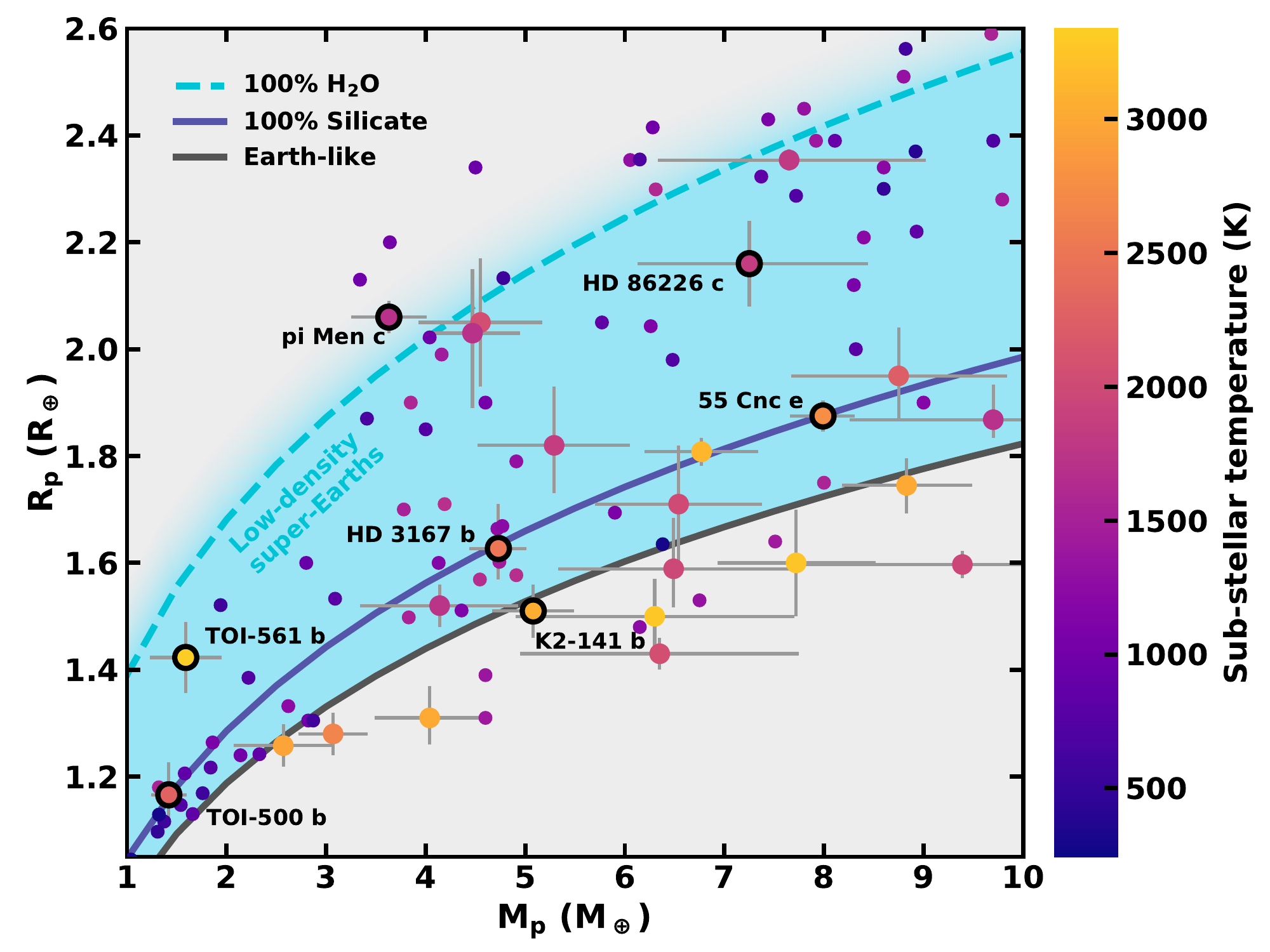}
    \includegraphics[width=0.49\textwidth]{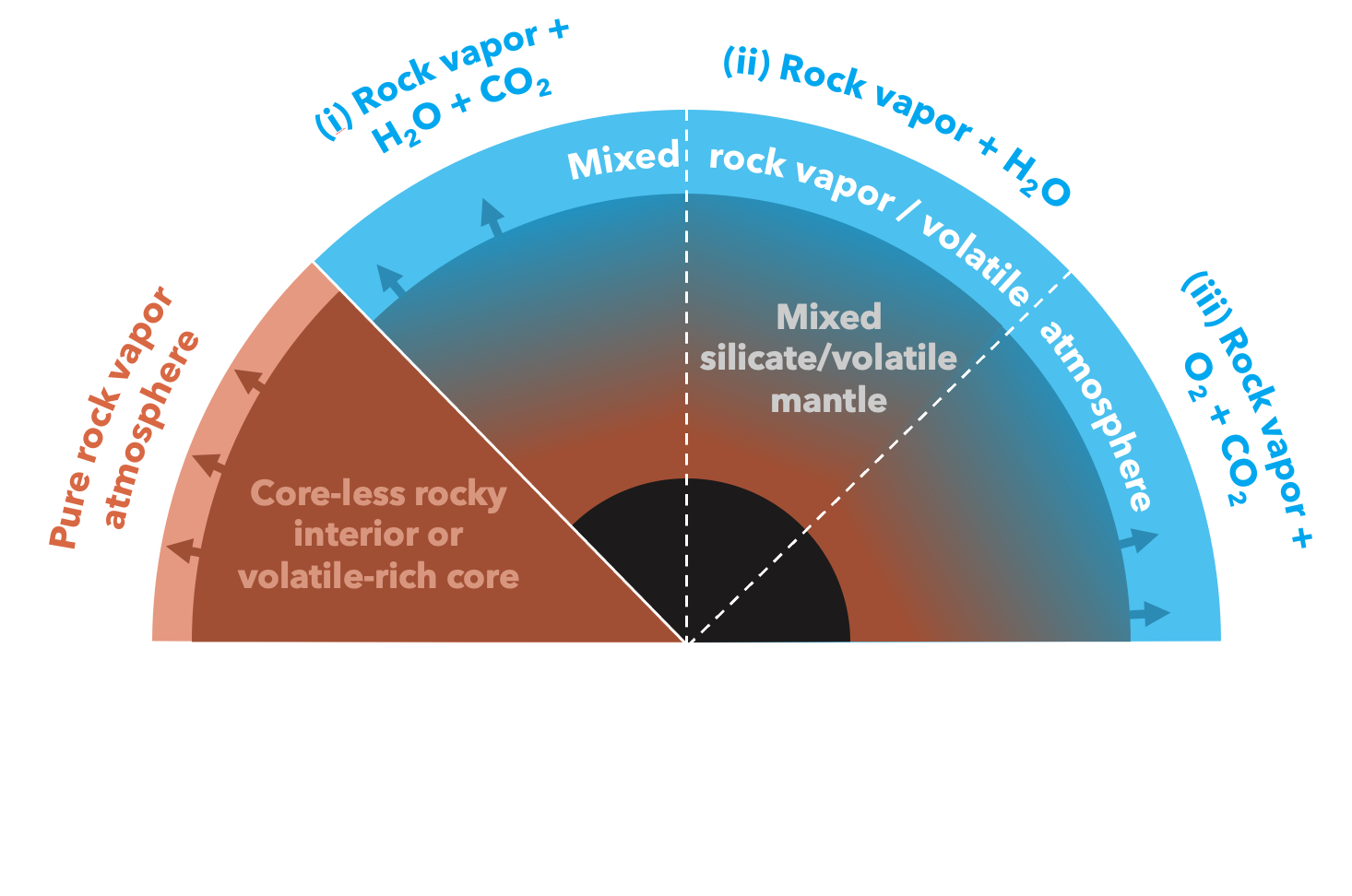}
    \caption{Left: A mass-radius ($M$-$R$) diagram showing exoplanets with masses and radii known to within 25\%. Planets in the blue shaded region are less dense than an Earth-like interior composition, but do not require a hydrogen-rich atmosphere to explain their radii (``low-density super-Earths''). The upper boundary of this region is blurred to represent the temperature dependence of mass-radius curves for water-rich planets. Lines show $M$-$R$ relations for an Earth-like (solid grey), 100\% MgSiO$_3$ (solid purple) and 100\% H$_2$O composition ($\sim$300~K, dashed blue), from \citet{Madhusudhan2020}. The color scale shows the sub-stellar temperature for each planet. Larger points with error bars show planets with sub-stellar temperatures $>1700$~K. For clarity, error bars are not shown for the smaller points (sub-stellar temperature $<1700$~K). Notable targets with high Emission Spectroscopy Metrics (ESMs, \citealt{Kempton2018}) are labeled. Mass and radius data are from the NASA Exoplanet Archive. Right: The atmospheric scenarios considered in this work for low-density lava worlds.}
    \label{fig:MR}
\end{figure*}

Efforts to characterize the masses and radii of close-in, low-mass exoplanets ($\lesssim$10~M$_\oplus$) have uncovered a diversity of planetary bulk densities \citep{Rogers2015,Zeng2021,Luque2022}. The radii of such planets follow a bimodal distribution, with peaks at $\sim$1.3~R$_\oplus$ and $\sim$2.4~R$_\oplus$ which can be explained by smaller, rocky planets and larger planets with hydrogen-rich envelopes, respectively \citep{Fulton2017,Owen2017,Ginzburg2018,Mordasini2020}. Meanwhile, mass measurements have revealed a growing population of planets with intermediate bulk densities \citep[e.g.,][]{Zeng2021,Luque2022,DiamondLowe2022,Brinkman2022,Cadieux2022,Piaulet2022}, which we refer to as low-density super-Earths (shaded blue region in Figure \ref{fig:MR}). The radii of these planets are too large to be explained by a rocky, Earth-like interior composition, but small enough that they do not necessarily require a hydrogen-rich envelope to explain their low bulk density. Instead, these planets may be explained by a range of interior and atmospheric compositions, including volatile-rich interiors \citep{Zeng2021,Dorn2021,Luque2022,Schlichting2022}, core-less silicate interiors \citep{Elkins-Tanton2008,Lichtenberg2021}, and thin hydrogen envelopes \citep{Rogers2023} -- or a combination of these. In order to break this degeneracy, atmospheric observations are needed to probe the chemical compositions of these planets.

While the atmospheric compositions of many super-Earths may be explained by degenerate mechanisms and interior compositions,
extremely irradiated super-Earths with evaporating surfaces can provide a direct probe of their interior compositions. Among the emerging population of low-density super-Earths, several planets have dayside temperatures exceeding the silicate melting point ($\sim$1700~K) -- ``low-density lava worlds''. The sub-stellar temperature of a planet is defined as
\begin{equation}
    \label{eq:Tsub}
    T_{\rm sub} = \sqrt{\frac{R_\star}{a}}T_{\rm eff},
\end{equation}
where $R_\star$ is the stellar radius, $a$ is the semi-major axis of the planet and $T_{\rm eff}$ is the effective temperature of the star. Notable low-density lava worlds with favorable observability metrics and high sub-stellar temperatures include HD~86226~c ($T_{\rm sub}$=1854~K, \citealt{Teske2020}), HD~3167~b ($T_{\rm sub}$=2513~K, \citealt{Vanderburg2016,Bourrier2022}), 55~Cnc~e ($T_{\rm sub}$=2773~K, \citealt{Bourrier2018,McArthur2004}) and TOI-561~b ($T_{\rm sub}$=3218~K, \citealt{Lacedelli2021,Weiss2021,Brinkman2022}), shown in Figure \ref{fig:MR}. Given their extremely high irradiation, the dayside atmospheres of these planets are expected to consist of evaporated surface material. In the case of a purely rocky surface, the atmosphere would consist of rock vapor species such as SiO, MgO, Na and K \citep{Ito2015,Miguel2011,Zilinskas2022}. In the case of a volatile-rich interior, the dayside atmosphere may additionally consist of volatile species evaporating from the surface and/or outgassing from the interior. Atmospheric observations of these planets therefore provide a unique opportunity to probe their interior compositions.

Low-density lava worlds are not expected to host primordially-accreted, hydrogen-rich atmospheres due to their extremely high irradiation levels \citep[e.g.,][]{Bourrier2022}; their low bulk densities must therefore be a result of their interior compositions. For some of these planets (e.g., those on the 100\% silicate mass-radius curve, see Figure \ref{fig:MR}), a volatile-rich iron core \citep{Schlichting2022,Li_2019_bookchapter} or the absence of a significant iron core \citep{Elkins-Tanton2008,Lichtenberg2021} are sufficient to explain the low bulk density. More generally, the bulk densities of low-density lava worlds may be explained by a volatile-rich mantle \citep{Madhusudhan2012,Dorn2021}. In this scenario, volatiles may evaporate from the surface and be detectable in the atmosphere. Meanwhile, if pure rock vapor atmospheres are detected around low-density lava worlds, this may imply a volatile-rich or absent core, with implications for the formation of such planets.

Several studies have explored how volatiles can be stored in and above a rocky mantle and/or magma ocean. For example, \citet{Dorn2021} show that significant amounts of H$_2$O can be stored in a magma ocean thanks to its high solubility (see also \citealt{Young2023} and references therein). They find that the bulk density of 55~Cnc~e can be explained by a 5\% mass fraction of H$_2$O, partitioned between a wet magma ocean and a steam atmosphere. H$_2$O can also be stored in a solid mantle, though in smaller quantities. For example, \citet{Guimond2023} consider the water storage capacity of solid super-Earths as a function of planetary mass and mantle composition, finding maximum water capacities of $\mathcal{O}$(0.1\%) by mass. At high pressures and temperatures, H$_2$O-rock miscibility can allow much higher mass fractions of H$_2$O to be stored in the mantles of low-density lava worlds \citep{Kovacevic2022,Vazan2022}. In particular, \citet{Vazan2022} find that rock and H$_2$O are fully miscible at pressures as low as 100~bar for temperatures $\gtrsim2500$~K. This suggests that a lava world with a surface temperature $\gtrsim$2500~K could potentially host a mixed H$_2$O-rock surface beneath its atmosphere. 

Whether volatiles are miscible or dissolved in a planet's mantle/magma ocean, the high dayside temperatures of low-density lava worlds would cause these volatiles to evaporate into the atmosphere and escape to space over time \citep[e.g.,][]{Kasting1983,Luger2015,Guo2019,Johnstone2020}. Therefore, detecting volatiles in the atmosphere of a low-density lava world would suggest the presence of a significant volatile inventory, which has survived atmospheric escape over the lifetime of the planet.

The high dayside temperatures and short orbital periods of lava worlds make them amenable to atmospheric characterization. Indeed, several studies have investigated the chemistry and spectral appearance of lava world atmospheres, both for purely rocky lava worlds \citep[e.g.,][]{Schaefer2009,Miguel2011,Ito2015,Kite2016,Nguyen2020,Nguyen2022,Zilinskas2022} and magma oceans in contact with a primary atmosphere \citep[e.g.,][]{Kite2021,Misener2022,Misener2023,Zilinskas2023}. Observations of 55~Cnc~e have already provided some initial insights into its possible atmospheric conditions \citep{Ehrenreich2012,Demory2016,Angelo2017,Esteves2017,Jindal2020,Zhang2021,Keles2022,Mercier2022}. Furthermore, the high observed C/O ratio of the host star (1.12$\pm$0.19 compared to the solar value of 0.55, \citealt{Asplund2009,DelgadoMena2010,Madhusudhan2012}), combined with the low bulk density of 55~Cnc~e, has led to the suggestion of a carbon-rich interior for this planet \citep{Madhusudhan2012}. Upcoming JWST observations of 55~Cnc~e and other lava worlds (K2-141~b, GJ~367~b, K2-22~b, WASP-47~e and TOI-561~b) will provide new insights into their atmospheric and surface conditions (Cycle 1 PIs and proposal numbers: R. Hu \#1952, A. Brandeker \#2084, N. Espinoza \#2159, L. Dang \#2347, M. Zhang \#2508; Cycle 2: J. Wright \#3315, S. Zieba \#3615, J. Teske \#3860). %\citep{Brandeker2021_jwstprop, Dang2021_jwstprop,Espinoza2021_jwstprop,Hu2021_jwstprop,Zhang2021_jwstprop}.

In this work, we investigate the observability of low-density lava world atmospheres with rocky to volatile-rich compositions (right panel of Figure \ref{fig:MR}). Observations of such atmospheres probe the composition of evaporated surface material, and therefore offer a unique observational window into their interiors. Using self-consistent atmospheric models, we study the impact of volatiles on the temperature structures and thermal emission spectra of low-density lava worlds, finding scenarios in which volatile spectral features may be observable and therefore indicate the presence of volatiles at the surface. The presence or absence of such signatures could provide important constraints on whether volatile-rich interiors are responsible for the low bulk densities of some super-Earths.

We describe our modeling framework in Section \ref{sec:methods}, including the opacity sources expected to dominate in mixed volatile/rock vapor atmospheres. In Section \ref{sec:target_selection}, we identify optimal low-density lava world targets for atmospheric characterization. Focusing on three such case studies spanning a range of sub-stellar temperatures ($\sim$1900--2800~K), we model the temperature structures and thermal emission spectra expected for a range of rock vapor/volatile ratios and for different volatile species in Section \ref{sec:results}. We assess the observability of various spectral features by simulating JWST observations for these case studies, and quantify the observing time needed to detect these features using atmospheric retrievals. Finally, we discuss our results and summarize our conclusions in sections \ref{sec:discussion} and \ref{sec:conclusions}, respectively.

\section{Methods}
\label{sec:methods}

\begin{figure*}
    \centering
    \includegraphics[width=0.8\textwidth]{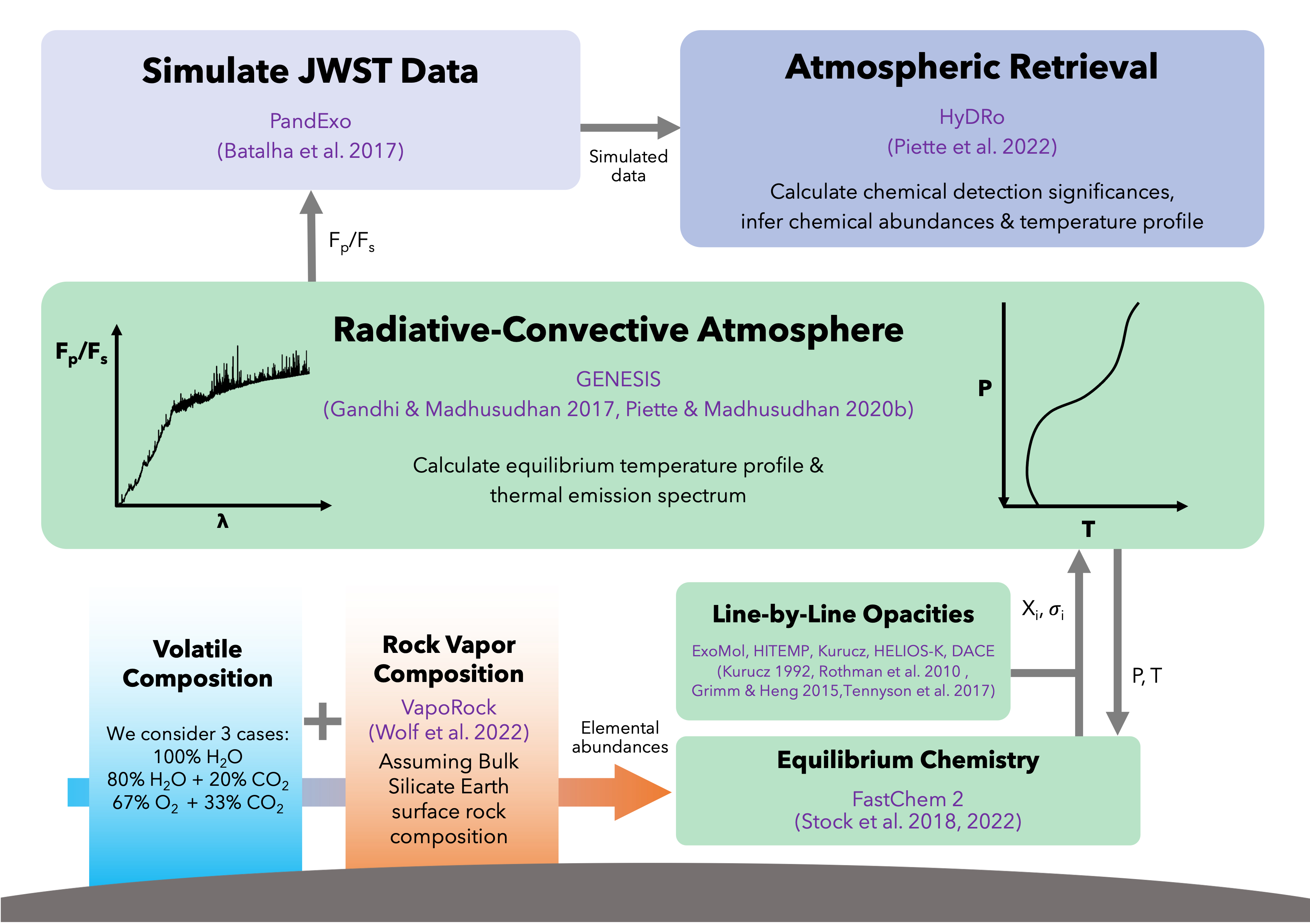}
    \caption{A schematic of the model framework used. We consider mixed atmospheric compositions consisting of volatiles and rock vapor. The atmospheric temperature profile and thermal emission spectrum of the planetary dayside are then calculated under the assumptions of radiative-convective, thermochemical and hydrostatic equilibrium. We simulate JWST observations corresponding to the model thermal emission spectrum and perform atmospheric retrievals to quantify the observability of key chemical species.}
    \label{fig:model}
\end{figure*}

In this work, we investigate the observability of low-density lava worlds using self-consistent atmospheric models across a range of conditions, from volatile-poor to volatile-rich. Here, we describe our modeling framework, shown schematically in Figure \ref{fig:model}. In order to self-consistently model the atmospheric structure, chemistry and thermal emission spectrum, we couple the \texttt{GENESIS} 1D atmospheric model \citep{Gandhi2017,Piette2020_MN} to the \texttt{FastChem 2} chemical equilibrium code \citep{Stock2018,Stock2022} and the \texttt{VapoRock} magma ocean outgassing code \citep{Wolf2022}. We begin by investigating the opacity structures of low-density lava worlds in Section \ref{sec:opacity}. We then describe the self-consistent atmospheric models in Section \ref{sec:forward_model}. In order to assess the observability of spectral features in the modeled planetary spectra, we simulate JWST observations using \texttt{Pandexo} \citep{Batalha2017} and quantify molecular detection significances using the \hydro atmospheric retrieval framework \citep{Piette2022} (Section \ref{sec:methods:retrievals}).

\subsection{Atmospheric composition and opacity}
\label{sec:opacity}

The bulk densities of low-density lava worlds may be explained by the presence of volatile species in or above their silicate mantles, as discussed in Section \ref{sec:intro}. We therefore model their atmospheres assuming a mixed rock-vapor/volatile composition, to account for the possibility of volatiles and/or silicates evaporating from the surface. The relative abundances of these volatile and rock vapor components depend on a number of unconstrained processes, including the solubility/miscibility of the volatiles in the mantle, the efficiency of mantle mixing processes, and the relative efficiencies of atmospheric loss for different chemical species. To marginalise over these processes, we consider a range of volatile/rock vapor ratios, from rock vapor-dominated to volatile-dominated atmospheres. In particular, in Section \ref{sec:results}, we show models spanning rock vapor fractions of 1 -- 100\%, where the remainder of the atmosphere consists of volatiles.

We consider three different compositions for the volatile component of the model atmosphere: 100\% H$_2$O, 80\% H$_2$O $+$ 20\% CO$_2$ and 67\% O$_2$ $+$ 33\% CO$_2$ (Figure \ref{fig:MR}). The 80\% H$_2$O $+$ 20\% CO$_2$ composition is motivated by average cometary compositions \citep{Ootsubo2012,Dello_Russo2016,McKay2019}, while the 100\% H$_2$O case explores an end-member scenario in which carbon species are not present at the planetary surface. Since water photolysis and loss of hydrogen to space is predicted to lead to the build-up of atmospheric O$_2$ in some cases \citep{Wordsworth2014,Luger2015}, we also consider the 67\% O$_2$ $+$ 33\% CO$_2$ composition to explore the impact of rapid hydrogen loss on the observability of volatiles.

We calculate the composition of the rock vapor component using the \texttt{VapoRock} magma ocean outgassing code \citep{Wolf2022}. The inputs of \texttt{VapoRock} are the surface composition, pressure, temperature and oxygen fugacity. \texttt{VapoRock} then outputs the partial pressures of each of the outgassed species. In our atmospheric model, these partial pressures are used to determine the relative abundances of rock vapor species in the atmosphere. We assume the surface composition to be that of the Bulk Silicate Earth, i.e., 45.97\% SiO$_2$, 36.66\% MgO, 8.24\% FeO, 4.77\% Al$_2$O$_3$, 3.78\% CaO, 0.35\% Na$_2$O, 0.18\% TiO$_2$ and 0.04\% K$_2$O \citep{Schaefer2009,Jackson2000}. For the oxygen fugacity, we use a nominal log$f$O$_2$ value of $\Delta$IW=+1.5\footnote{i.e., a 1.5 dex deviation in oxygen fugacity compared to the Iron-W\"ustite buffer, see \citet{ONeill1993,Hirschmann2008,Wolf2022}}, following \citet{Wolf2022}. We find that the input surface pressure has a negligible effect on the output abundances for values $\lesssim 100$~bar (see also \citealt{vanBuchem2022}), so we use a fixed surface pressure of 1~mbar for the \texttt{VapoRock} calculations. The input surface temperature used by \texttt{VapoRock} can be set independently from the equilibrium temperature profile calculated by the radiative-convective atmospheric model (described in Section \ref{sec:forward_model}). We find that differences in the \texttt{VapoRock} input surface temperature have little effect on the resulting temperature profile and emission spectrum of the atmosphere (Appendix \ref{sec:appendix:Tsurf}). For simplicity, we therefore set the \texttt{VapoRock} input surface temperature equal to the sub-stellar temperature of the planet. This temperature is typically close to the temperature at the base of the photosphere in our models. 

We note that for a dry magma ocean, the surface composition may not be representative of the bulk mantle \citep[e.g.][]{Kite2016,Boukare2022}. For example, rapid vaporization of the most volatile species can result in a chemically distilled surface composition which is not representative of the bulk interior \citep{Kite2016,Nguyen2020,Zilinskas2022}. For simplicity, we do not consider such effects in this work and instead focus on the effects of adding volatile species.

Given a particular volatile/rock vapor ratio, the volatile composition and the outgassed rock vapor composition, we calculate the equilibrium chemical abundances in the atmosphere using \texttt{FastChem 2} \citep{Stock2018,Stock2022}. We input the elemental abundances based on the compositions of the volatile and rock vapor components, and calculate equilibrium abundances as a function of pressure and temperature in the atmosphere. The atmospheric temperature profile and chemical profile are calculated iteratively by coupling \texttt{FastChem 2} with the self-consistent atmospheric model \texttt{GENESIS}  \citep{Gandhi2017}, as described in Section \ref{sec:forward_model}.

The atmospheric composition is linked to the equilibrium temperature profile through the opacities of the chemical species present. For the calculation of the equilibrium temperature profile, we include UV to infrared opacity from the key chemical species in the atmosphere. For the molecular species, we calculate absorption cross sections using the methods described in \citet{Gandhi2017} and line list data from the ExoMol, HITEMP and HITRAN databases \citep{Rothman2010,Tennyson2016,Gordon2017}. In particular, we consider opacity from the following molecular species and corresponding line lists: H$_2$O \citep{Rothman2010}, CO$_2$ \citep{Rothman2010}, CO \citep{Rothman2010}, SiO \citep{Yurchenko2021}, SiO$_2$ \citep{Owens2020}, AlO \citep{Patrascu2015}, MgO \citep{Li2019}, NaO \citep{Mitev2022}, TiO \citep{McKemmish2019}, O$_2$ \citep{Gordon2017}, OH \citep{Rothman2010}, FeH \citep{Dulick2003,Bernath2020}, NaH \citep{Rivlin2015}, NaOH \citep{Owens2021} and KOH \citep{Owens2021}. For the atomic and ionic species, we use opacities from the \texttt{DACE}\footnote{https://dace.unige.ch/} database \citep{Grimm2021}, calculated using \texttt{helios-k} \citep{Grimm2015,Grimm2021} and data from the Kurucz\footnote{http://kurucz.harvard.edu/} database \citep{Kurucz1992,Kurucz2017,Kurucz2018}. We consider opacity from the following atomic and ionic species: Al, Ca, Fe, H, K, Mg, Na, O, Si, Ti, Ca$^+$, and Na$^+$.

\begin{figure*}
    \centering
    \includegraphics[width=1\textwidth]{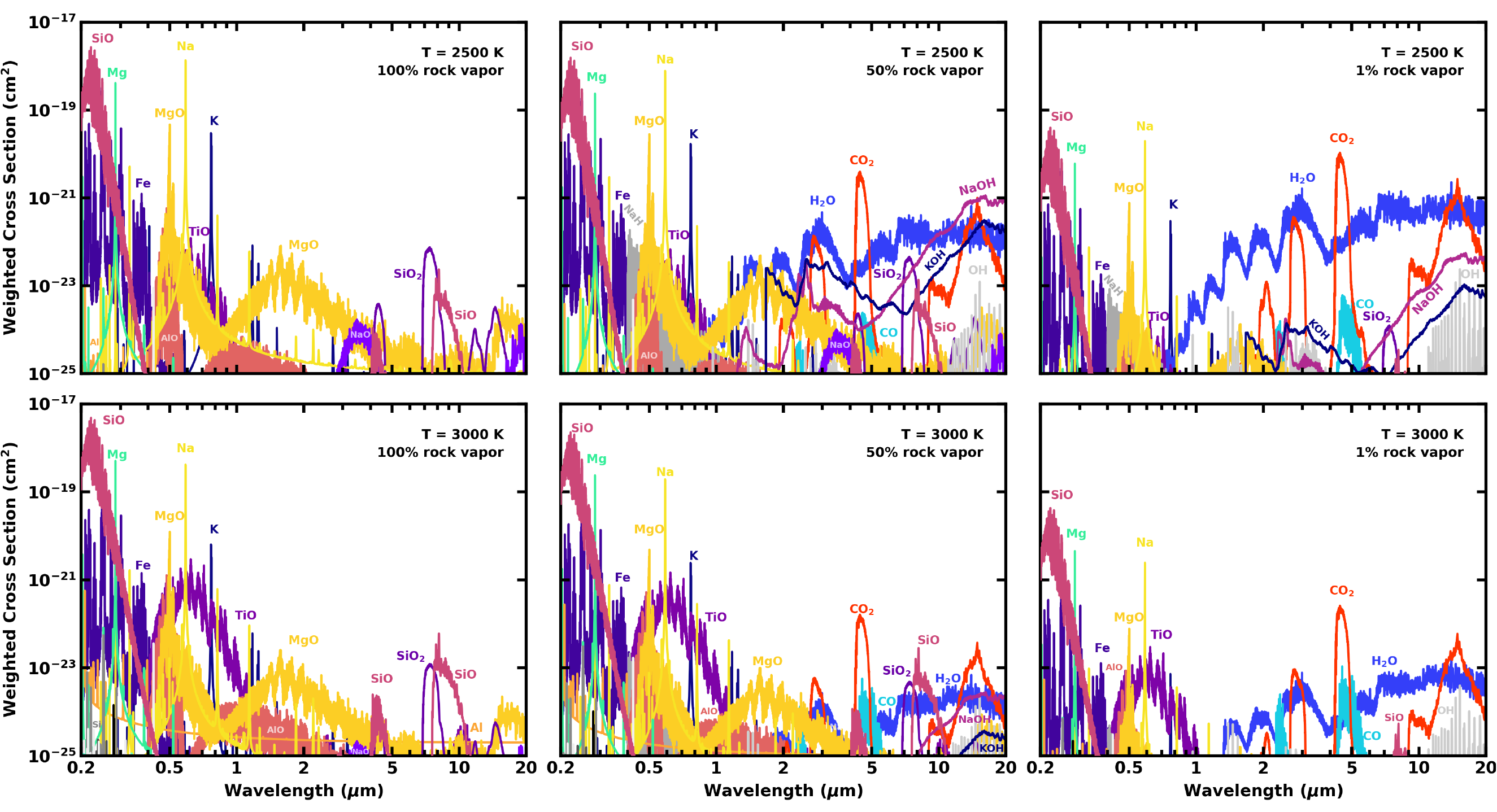}
    \caption{Absorption cross sections weighted by volume mixing ratio. Panels from left to right show the dominant opacity sources for three atmospheric compositions, which have elemental abundances corresponding to: 100\% rock vapor (left) 50\% rock vapor + 40\% H$_2$O + 10\% CO$_2$ (middle) and 1\% rock vapor + 79\% H$_2$O + 10\% CO$_2$ (right). Top and bottom rows show weighted cross sections corresponding to a temperature of 2500~K and 3000~K, respectively, and a pressure of 1~mbar. }
    \label{fig:opac}
\end{figure*}

Figure \ref{fig:opac} shows abundance-weighted absorption cross sections in the 0.2--20~$\mu$m range for the dominant chemical species given three atmospheric compositions: 100\% rock vapor, 50\% rock vapor + 50\% volatiles, and 1\% rock vapor + 99\% volatiles. Here, the volatile component is assumed to be 80\% H$_2$O + 20\% CO$_2$ and the pressure is fixed to 1~mbar. The top/bottom three panels show weighted cross sections at 2500~K/3000~K, respectively. Across all three compositions, the UV and optical opacity is dominated by rock vapor species, notably from SiO and Fe in the UV, and TiO, MgO, AlO, Na and K in the optical. For a pure rock vapor atmosphere, MgO acts as a source of continuum opacity in the infrared \citep{Zilinskas2022}, while SiO and SiO$_2$ have strong infrared features in the $\sim$7-11~$\mu$m range.

The addition of volatiles to the atmosphere increases the overall infrared opacity. In particular, H$_2$O, CO$_2$ and CO contribute significant infrared opacity, though H$_2$O and CO$_2$ are thermally dissociated at higher temperatures. For intermediate compositions with significant contributions from both volatiles and rock vapor, the metal hydroxides NaOH and KOH are present in chemical equilibrium, and contribute strong infrared opacity at wavelengths $\gtrsim$10~$\mu$m. As the volatile/rock vapor ratio increases, the balance of infrared to UV/optical opacity also increases, resulting in temperature profiles that have weaker thermal inversions, or which are non-inverted \citep[e.g.,][]{Hubeny2003}. Furthermore, H$_2$O, CO$_2$ and CO all have strong spectral features in the infrared wavelength range probed by JWST, and may therefore be detectable in the atmospheres of low-density lava worlds. In particular, CO and CO$_2$ have strong spectral features in the $\sim$4-5~$\mu$m range, while H$_2$O has multiple spectral features across the infrared.

\subsection{Self-consistent 1D atmospheric model}
\label{sec:forward_model}

We use the \texttt{GENESIS} self-consistent 1D atmospheric model \citep{Gandhi2017,Piette2020_tio,Piette2020_MN} to calculate the equilibrium temperature profiles and corresponding thermal emission spectra of the low-density lava worlds considered in this work. \texttt{GENESIS} solves radiative-convective equilibrium under the assumptions of hydrostatic, thermochemical and local thermodynamic equilibrium. Equilibrium chemical abundances are calculated iteratively alongside the temperature profile by coupling \texttt{GENESIS} to \texttt{FastChem 2}. The energetic boundary conditions of the atmosphere are set by the stellar irradiation and the internal heat emanating from the centre of the planet. We use Kurucz stellar spectra \citep{Kurucz1979,Castelli2003} to model the incident stellar irradiation, and assume an Earth-like internal temperature of 10~K. We note that, given the high irradiation temperatures we consider, the internal temperature has a negligible effect on the dayside atmosphere. 

The dayside temperature profile also depends on the efficiency of energy redistribution in the atmosphere. This can be characterised by the redistribution parameter, $f$, such that
\begin{equation}
    T_\mathrm{day} = f^{1/4}(1-A_\mathrm{B})^{1/4}\sqrt{\frac{R_\star}{a}}T_\mathrm{eff},
\end{equation}
where $A_\mathrm{B}$ is the Bond albedo, $f=0.25$ in the case of efficient day-night energy redistribution, and $f=2/3$ in the case of no energy redistribution (i.e., instant re-radiation, see \citealt{Burrows2008}). High atmospheric temperatures typically reduce the efficiency of energy redistribution as a result of very short radiative timescales and/or frictional drag due to Lorentz forces in a partially ionized atmosphere \citep{Komacek2016}. Therefore, given the high temperature regime explored in this work, we assume $f=2/3$ in our models.

The total atmospheric pressure of a lava world depends on the competing rates of surface outgassing and atmospheric escape. For purely rocky lava worlds, the extent of surface outgassing depends on temperature, where higher surface temperatures lead to increased outgassing and a higher atmospheric pressure \citep{Miguel2011,Ito2015,Zilinskas2022}. Furthermore, \citet{Ito2021} show that atmospheric loss is reduced for rock vapor atmospheres due to atomic line cooling. Meanwhile, lava worlds with volatile-rich interiors may have significantly higher surface pressures due to redistribution of volatiles between the atmosphere and molten interior. In this work, we assume that atmospheric escape is a relatively small effect compared to outgassing from the interior due to the rapid convection in magma oceans \citep{Lichtenberg2022,Salvador2023}. 

For pure rock vapor atmospheres, we use the total atmospheric pressure output by \texttt{VapoRock} (see Section \ref{sec:opacity}), which for the cases shown corresponds to atmospheres which are close to being optically thick. However, for compositions including volatiles, the partial pressure of outgassed volatiles is not known a priori, and may depend on whether the volatiles are dissolved in a magma ocean or exist as a super-critical fluid. We therefore make the simplifying assumption that the atmosphere is optically thick, meaning that the emergent spectrum is not sensitive to pressures deeper than the photosphere. In figures \ref{fig:HD86226c_results}--\ref{fig:55Cnce_results}, temperature profiles are shown from the base of the 0.2-50~$\mu$m photosphere down to a pressure of 10$^{-6}$~bar. We do not include the effects of a surface in our models, as such effects would only be relevant for optically thin atmospheres.

Given the energetic constraints and pressure boundaries described above, the equilibrium temperature profile and thermal emission spectrum are solved for iteratively using the Rybicki scheme \citep{Hubeny2014, Gandhi2017}. This calculation requires the radiative transfer equation to be solved in each iteration; we do this using the Feautrier scheme \citep{Feautrier1964}, which is fast and provides the required Eddington factors. Once the equilibrium temperature profile is found, the final emergent planetary spectrum is calculated using a combination of the Discrete Finite Element method \citep{Castor1992} and Accelerated Lambda Iteration \citep{Hubeny2014,Hubeny2017}. This method provides a more accurate solution for the mean intensity, which is required for this calculation of the planetary flux \citep{Piette2020_MN}. Using the final thermal emission spectrum, we also simulate JWST observations to assess the observability of key spectral features, as described in Section \ref{sec:methods:retrievals}.

\subsection{Simulated data \& atmospheric retrievals}
\label{sec:methods:retrievals}

We simulate JWST observations corresponding to the model thermal emission spectra calculated using \texttt{GENESIS}. We focus on the $\sim$3-12~$\mu$m range, where the signal to noise is ideal and key spectral features are present due to H$_2$O, CO$_2$, SiO and SiO$_2$. In particular, we use \texttt{PandExo} \citep{Batalha2017} to simulate data for the NIRSpec~G395H, NIRCam~F444W and MIRI~LRS instrument modes. For each of these cases, we generate model spectra at a resolution of $R\sim$15,000, convolve to the instrument resolution using a Gaussian kernel, and then bin to the data resolution output by \texttt{PandExo}. We further calculate the data uncertainties using \texttt{PandExo}, assuming a noise floor of 5~ppm at native resolution \citep{Lustig-Yaeger2023}. These uncertainties are added to the simulated data as random Gaussian noise.

In order to statistically assess the observability of molecular features in the model emission spectra, we perform atmospheric retrievals using the \hydro retrieval framework \citep{Piette2022}. \hydro is built upon the \texttt{HyDRA} atmospheric retrieval framework \citep{Gandhi2018,Gandhi2020,Piette2020_BD}, and consists of a parametric forward atmospheric model coupled to a Nested Sampling Bayesian parameter estimation algorithm \citep{Skilling2006}, \texttt{PyMultiNest} \citep{Feroz2009,Buchner2014}. The free parameters to the forward model include the temperature profile parameters and the constant-with-depth abundances of the chemical species included. Here, we use the 6-parameter temperature profile parameterization of \citet{Madhusudhan2009}, which is able to capture the wide range of temperature structures expected in exoplanet atmospheres, from thermally inverted to non-inverted profiles. The prior probability distributions used for the temperature profile parameters are shown in Table \ref{tab:PTpriors}. 

\begin{table}
    \centering
    \caption{Prior probability distributions for the temperature profile parameters in the atmospheric retrievals shown in Section \ref{sec:results:observability}.}
    \begin{tabular}{c|c|c}
       \bf Parameter (units) &  \bf Prior Distribution & \bf Range\\
       \hline
       $\alpha_1$ (K$^{-1/2}$) & uniform & 0.02 $-$ 1 \\
       $\alpha_2$ (K$^{-1/2}$) & uniform & 0.02 $-$ 1\\
       $T_{100 \rm mb}$ (K) & uniform & 1000 $-$ 4000\\
       $P_1$ (bar) & log-uniform & $10^{-5}$ $-$ 100\\
       $P_2$ (bar) & log-uniform & $10^{-5}$ $-$ 100\\
       $P_3$ (bar) & log-uniform & $10^{-2}$ $-$ 100\\
    \end{tabular}
    \label{tab:PTpriors}
\end{table}

Unlike giant planet atmospheres, which are known to have H$_2$-rich compositions, the atmospheres of low-mass exoplanets can span a wide range of background chemical compositions. \hydro remains agnostic about the background composition of the atmosphere by using the Centred-Log-Ratio (CLR) parameterization for the chemical abundances \citep{Benneke2012,Piette2022}. This parameterization results in identical priors for the abundances of each of the chemical species in the retrieval. The priors we use for the CLR abundance parameters correspond to lower limits of 10$^{-15}$ on the mixing ratios for each chemical species. From the forward models described in sections \ref{sec:opacity} and \ref{sec:forward_model}, we find that H$_2$O, CO$_2$, CO, SiO, SiO$_2$ and MgO are the dominant infrared opacity sources across a range of rock vapor/volatile ratios (Figure \ref{fig:opac}). We therefore include these 6 species in the retrieval, using the absorption cross sections described in Section \ref{sec:opacity}.

We use the \hydro atmospheric retrievals to statistically quantify the observability of H$_2$O and CO$_2$ features in the atmospheres of low-density lava worlds. To do this, we perform model comparisons between retrievals which include/exclude a particular molecule. The ratio of the Bayesian evidences for each retrieval model (i.e., the Bayes' factor) is:
\begin{equation}
    \mathcal{B} = \frac{p(\mathrm{data}|\mathrm{model\, with\,molecule})}{p(\mathrm{data}|\mathrm{model\, without\,molecule})}.
\end{equation}
$\mathcal{B}=$1.0, 2.5 or 5.0 suggests weak, modest and strong evidence, respectively, for the model which includes the molecule in question. This Bayes' factor can, in turn, be converted to a `sigma' value to assess the confidence of the molecular detection \citep{Benneke2013}. In Section \ref{sec:results:observability}, we use this approach to assess the observing time required to achieve $\gtrsim 3\sigma$ detections of H$_2$O and CO$_2$.

\section{Ideal targets}
\label{sec:target_selection}

\begin{figure}
    \centering
    \includegraphics[width=0.5\textwidth]{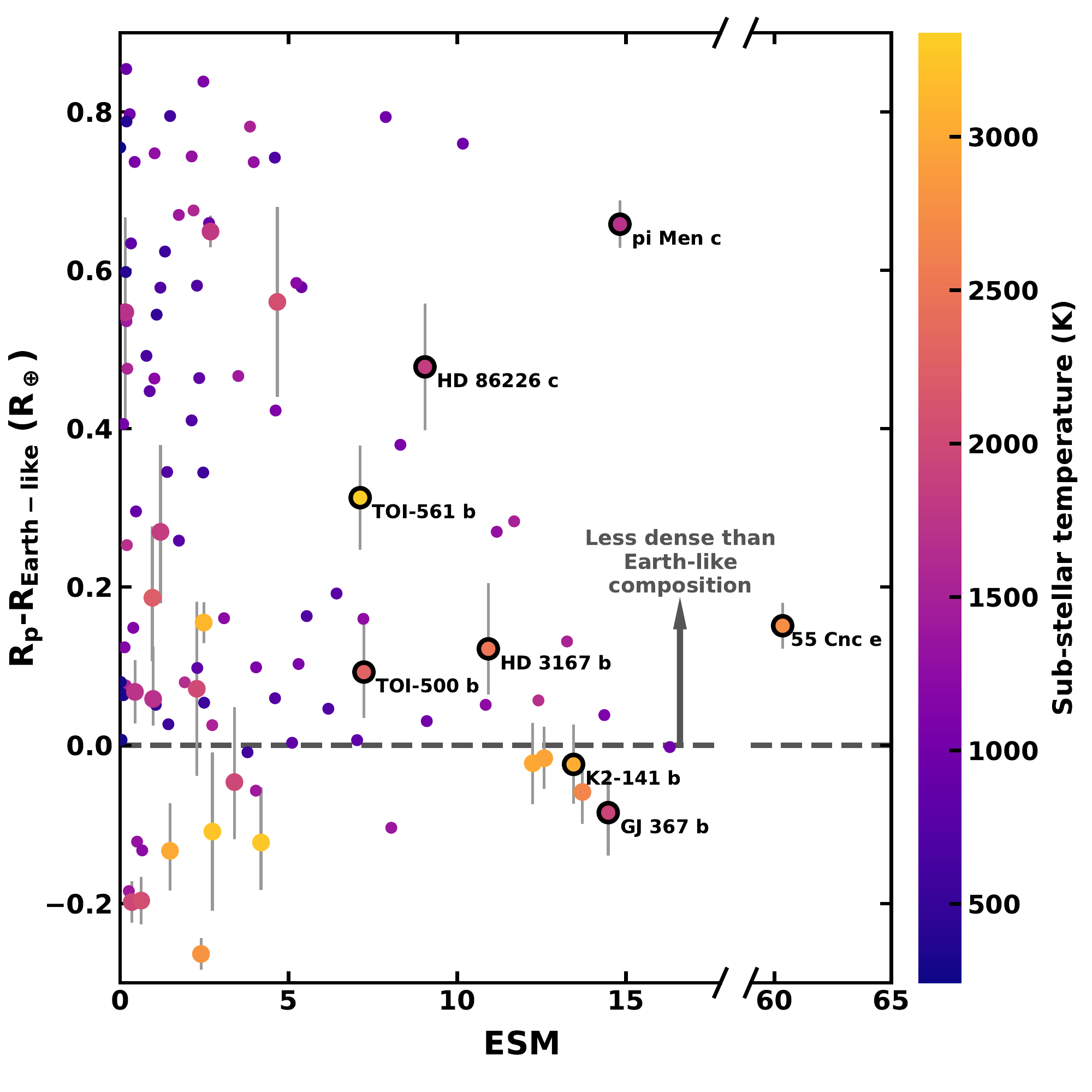}
    \caption{Planetary radius relative to the radius expected for an Earth-like interior composition, as a function of the Emission Spectroscopy Metric (ESM). The color scale shows the sub-stellar temperature for each planet. Larger points with error bars show planets with a sub-stellar temperature $>1700$~K, which are therefore hot enough for the dayside to be partially molten. Smaller points denote planets with sub-stellar temperatures cooler than 1700~K. The case studies explored in this work (HD~86226~c, HD~3167b and 55~Cnc~e), other lava worlds with ESM>5 (pi~Men~c, TOI-500~b, TOI-561~b), and planets scheduled for JWST Cycle 1 observations (GJ~367~b, K2-141~b and 55~Cnc~e) are labelled and outlined with bold black circles.}
    \label{fig:targets}
\end{figure}

\begin{table*}
    \centering
    \hspace{-4cm}
    \begin{tabular}{l|c|c|c|c|c|c|c|c|c}
        \bf Planet & $\bf M_\mathrm{\bf p}$ ($\bf M_\oplus$) & $\bf R_\mathrm{\bf p}$ ($\bf R_\oplus$) & \bf T$_\mathrm{\mathbf{sub}}$ (K) & $\bf a$ (au) & \bf T$_\mathrm{\bf eff}$ (K) & \bf log(g$_\star$/cgs) & $\bf R_\star$ ($\bf R_\odot$) & \bf K mag & \bf $\mathbf{d}$ (pc)\\
        \hline
        HD~86226~c & 7.25 & 2.16 & 1854 & 0.049 & 5863 & 4.40 & 1.053 & 6.463 & 45.6830 \\ 
        HD~3167~b & 4.73 & 1.627 & 2513 & 0.01802 & 5300 & 4.47 & 0.871 & 7.066 & 47.2899 \\
        55~Cnc~e  & 7.99 & 1.875 & 2773 & 0.01544 & 5172 & 4.43 & 0.943 & 4.015 & 12.5855  \\
    \end{tabular}
    \caption{Planetary and stellar parameters for the three case studies considered in this work: planet mass, radius, sub-stellar temperature (Equation \ref{eq:Tsub}) and semi-major axis, and host star effective temperature, gravity, radius, K magnitude and distance. Stellar K magnitudes are from 2MASS \citep{2MASS2003}. Other values for HD~86226~c, HD~3167~b and 55~Cnc~e are from \citet{Teske2020}, \citet{Bourrier2022} and \citet{Bourrier2018}, respectively.}
    \label{tab:planet_params}
\end{table*}

In this section, we assess the observability of known, low-density lava worlds in order to identify optimal targets for atmospheric observations with JWST. We consider known lava worlds with masses and radii constrained to within 25\% and with masses $<10$~M$_\oplus$ (data are from the NASA Exoplanet Archive\footnote{exoplanetarchive.ipac.caltech.edu}). We define a `lava world' as having a sub-stellar temperature $>1700$~K, i.e., exceeding the dry silicate melting point at low pressure. Figure \ref{fig:MR} shows that a number of these lava worlds have densities less than that expected for a 100\% silicate interior (solid purple line), and instead have bulk densities consistent with a volatile-enriched interior \citep{Dorn2021}, the lack of a significant iron core \citep{Elkins-Tanton2008,Lichtenberg2021} and/or a volatile-rich iron core \citep{Li_2019_bookchapter,Schlichting2022}. These targets include 55~Cnc~e \citep{McArthur2004,Bourrier2018}, HD~3167~b \citep{Vanderburg2016,Bourrier2022}, HD~86226~c \citep{Teske2020}, pi~Men~c \citep{Gandolfi2018,Huang2018,Hatzes2022}, TOI-500~b \citep{Giacalone2022,Serrano2022} and TOI-561~b \citep{Lacedelli2021,Weiss2021,Brinkman2022}, and span a wide range of masses from $\sim 1-8$~M$_\oplus$.

In addition to being consistent with possessing volatile-rich interiors, these low-density lava worlds have high Emission Spectroscopy Metrics (ESMs) \citep{Kempton2018}, making them ideal candidates for secondary eclipse observations with JWST. Figure \ref{fig:targets} shows the ESM in relation to the `excess' radius of each planet relative to the radius expected for an Earth-like interior. Lava worlds with larger excess radii have lower bulk densities, and may host volatile-rich interiors. Among the lava worlds shown here with a density lower than that expected for an Earth-like composition, the six lava worlds listed above have the highest ESMs. 

In what follows, we focus on three promising targets spanning sub-stellar temperatures from $\sim$1900--2800~K: HD~86226~c, HD~3167~b and 55~Cnc~e. In Section \ref{sec:results}, we explore the atmospheric structures and observability for these case studies across a range of possible volatile-rich to rock vapor-rich compositions. By spanning a range of irradiation temperatures, these case studies also demonstrate the effects of dayside temperature on the atmospheric chemistry and thermal emission spectra of such planets. The planetary and stellar parameters we use for these systems are shown in Table \ref{tab:planet_params}.

HD~86226~c \citep{Teske2020} is the coolest case study we consider, with a sub-stellar temperature of 1854~K. With a mass of 7.25~M$_\oplus$ and a radius of 2.16~R$_\oplus$, the bulk density of this planet is significantly lower than that expected for a 100\% silicate composition (\citealt{Teske2020}, see also Figure \ref{fig:MR}). This could be explained by either a thin, escaping hydrogen-rich atmosphere, or a volatile-rich interior composition. In this work, we focus on the latter case and test whether volatiles evaporating from the surface could be detected in the atmosphere.

The second case study we consider is HD~3167~b \citep{Vanderburg2016,Christiansen2017,Bourrier2022}. With a mass of 4.73~M$_\oplus$ and a radius of 1.63~R$_\oplus$, HD~3167~b's bulk density is consistent with a 100\% silicate interior composition (Figure \ref{fig:MR}). This can be explained by a core-less silicate interior \citep{Elkins-Tanton2008,Lichtenberg2021}, a volatile-rich iron core \citep{Li_2019_bookchapter,Schlichting2022} and/or a volatile-rich mantle \citep{Dorn2021}. \citet{Bourrier2022} rule out a hydrogen-rich atmosphere for this planet using atmospheric evolution models. With a sub-stellar temperature of $\sim$2500~K, a purely rocky surface would be expected to vaporize and form an optically thick, $\sim$10~mbar atmosphere \citep{Zilinskas2022}. If HD~3167~b's surface also contains volatiles, the atmosphere may be thicker and have observable signatures of volatile species such as H$_2$O and CO$_2$.

The hottest case study we consider is 55~Cnc~e \citep{McArthur2004,Fischer2008,Dawson2010,Winn2011}, a 7.99~M$_\oplus$, 1.88~R$_\oplus$ super-Earth. Similarly to HD~3167~b, the bulk density of 55~Cnc~e is consistent with a core-less silicate interior, a volatile-rich core and/or a volatile-rich mantle (Figure \ref{fig:MR}). 55~Cnc~e has been the subject of numerous atmospheric studies. Spitzer phase curve observations at 4.5~$\mu$m initially revealed a significant eastwards phase offset \citep{Demory2016,Angelo2017}, suggesting the presence of a substantial atmosphere \citep{Hammond2017}. However, a recent reanalysis of the Spitzer data by \citet{Mercier2022} indicates a negligible phase offset and a stronger day-night temperature contrast, consistent with either a local dayside atmosphere or a global atmosphere with poor heat redistribution efficiency. 55~Cnc~e will be observed spectroscopically in JWST's Cycle 1 using NIRCam (total of 5 eclipses) and MIRI~LRS (one eclipse) (Cycle 1 proposals \#2084, PI: A. Brandeker, \citealt{Brandeker2021_jwstprop} and \#1952, PI: R. Hu, \citealt{Hu2021_jwstprop}). In Section \ref{sec:results}, we simulate NIRCam and MIRI~LRS observations for 55~Cnc~e in order to predict whether volatiles escaping from the interior could be detected using these Cycle 1 observations.

\section{Results}
\label{sec:results}

\begin{figure*}
    \centering
    \includegraphics[width=0.95\textwidth]{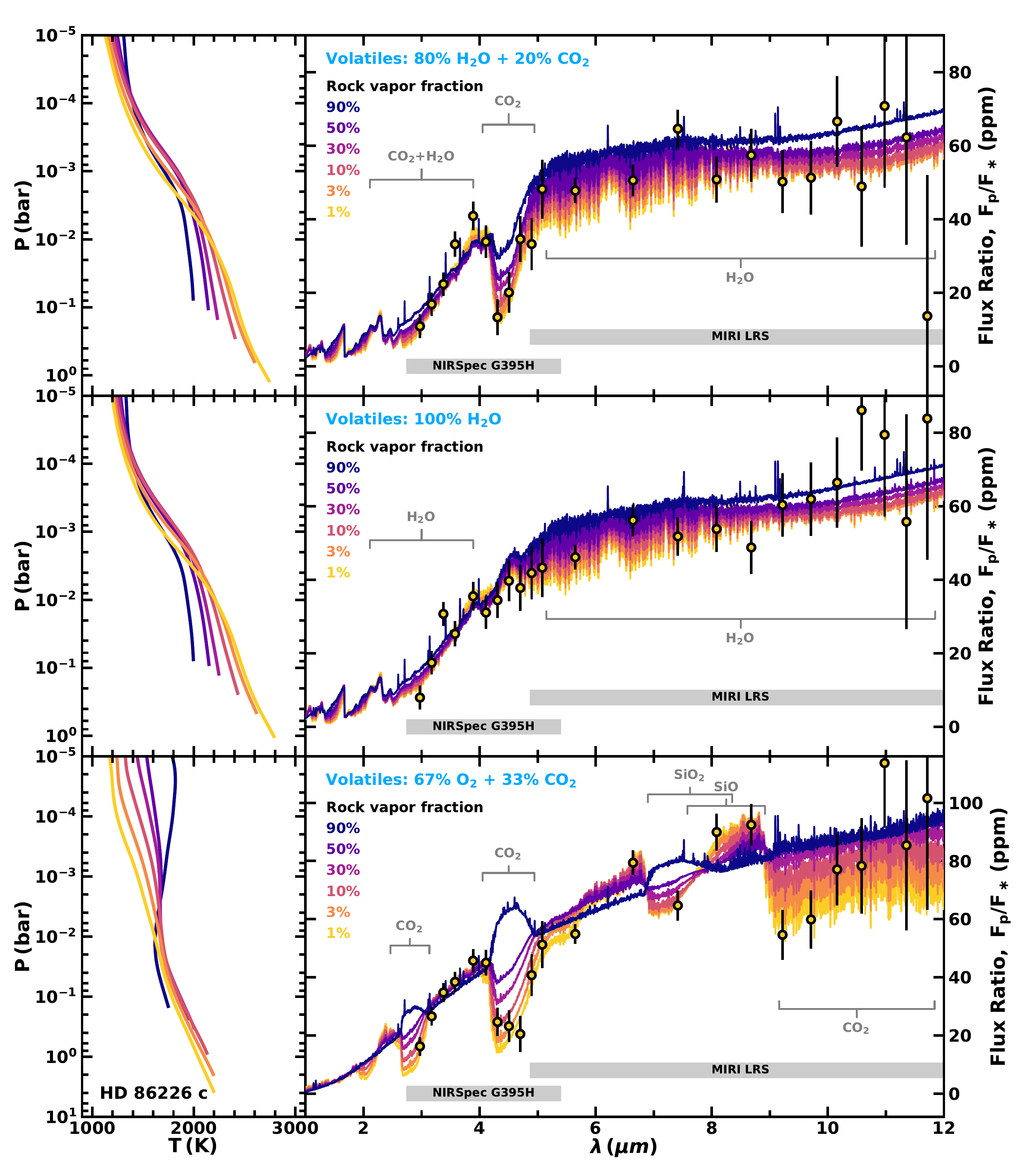}
    \caption{Model atmospheric temperature profiles (left) and thermal emission spectra (right) for HD~86226~c assuming mixed volatile/rock vapor compositions. In the top, middle and lower panels, the volatile component consists of 80\% H$_2$O $+$ 20\% CO$_2$, 100\% H$_2$O and 67\% O$_2$ $+$ 33\% CO$_2$, respectively. In each panel, lighter colors correspond to lower rock vapor fractions and higher volatile fractions. Simulated JWST observations are shown for the most volatile-rich cases assuming 5 NIRSpec~G395H eclipses and 5 MIRI~LRS eclipses. Temperature profiles are shown up to the pressure corresponding to the base of the 0.2 -- 50~$\mu$m photosphere. Key spectral features are labeled in grey.}
    \label{fig:HD86226c_results}
\end{figure*}

\begin{figure*}
    \centering
    \includegraphics[width=0.95\textwidth]{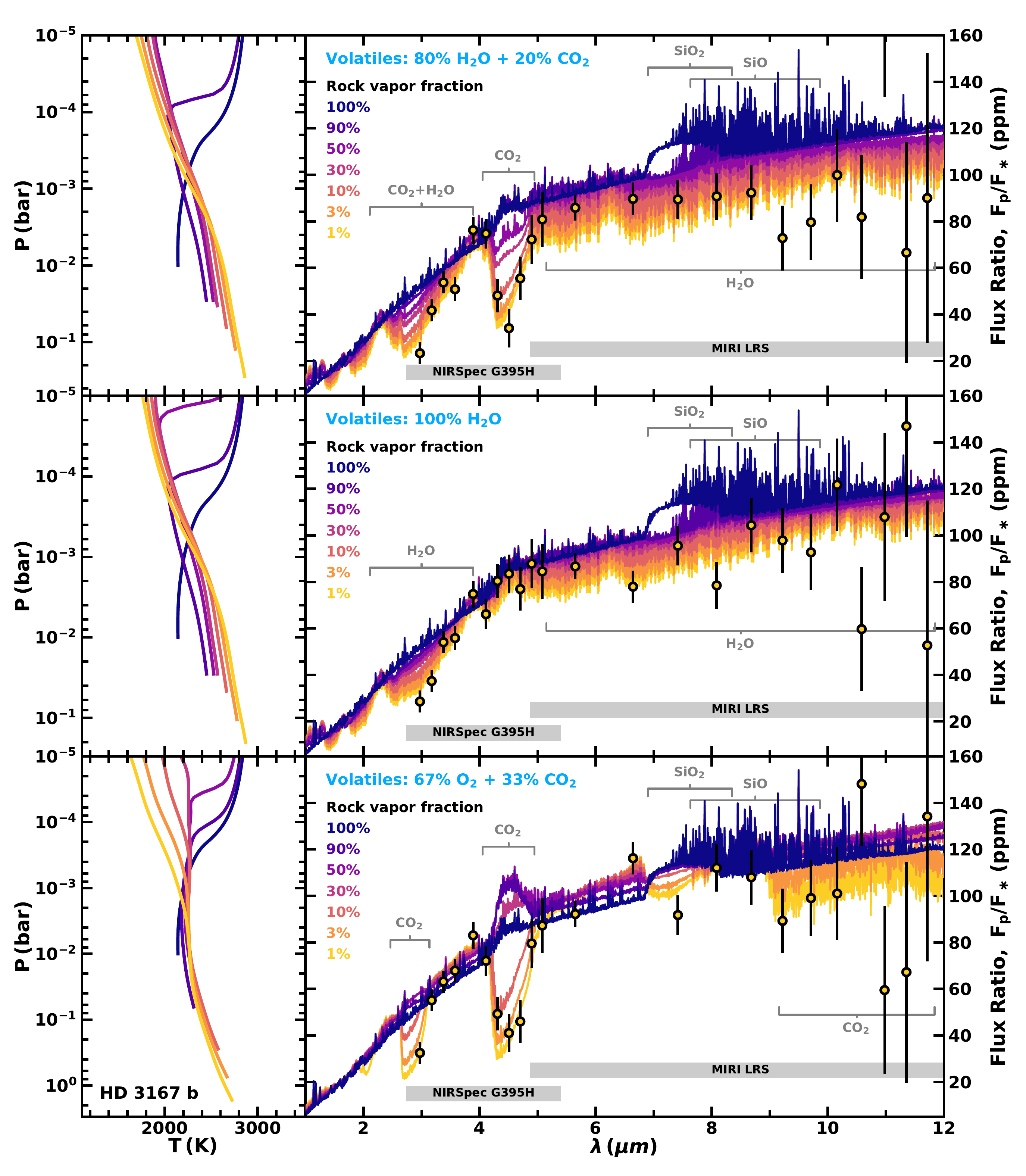}
    \caption{Same as Figure \ref{fig:HD86226c_results} but for HD~3167~b. Simulated JWST observations are shown for the most volatile-rich cases assuming 5 NIRSpec~G395H eclipses and 5 MIRI~LRS eclipses.}
    \label{fig:HD3167b_results}
\end{figure*}

\begin{figure*}
    \centering
    \includegraphics[width=0.95\textwidth]{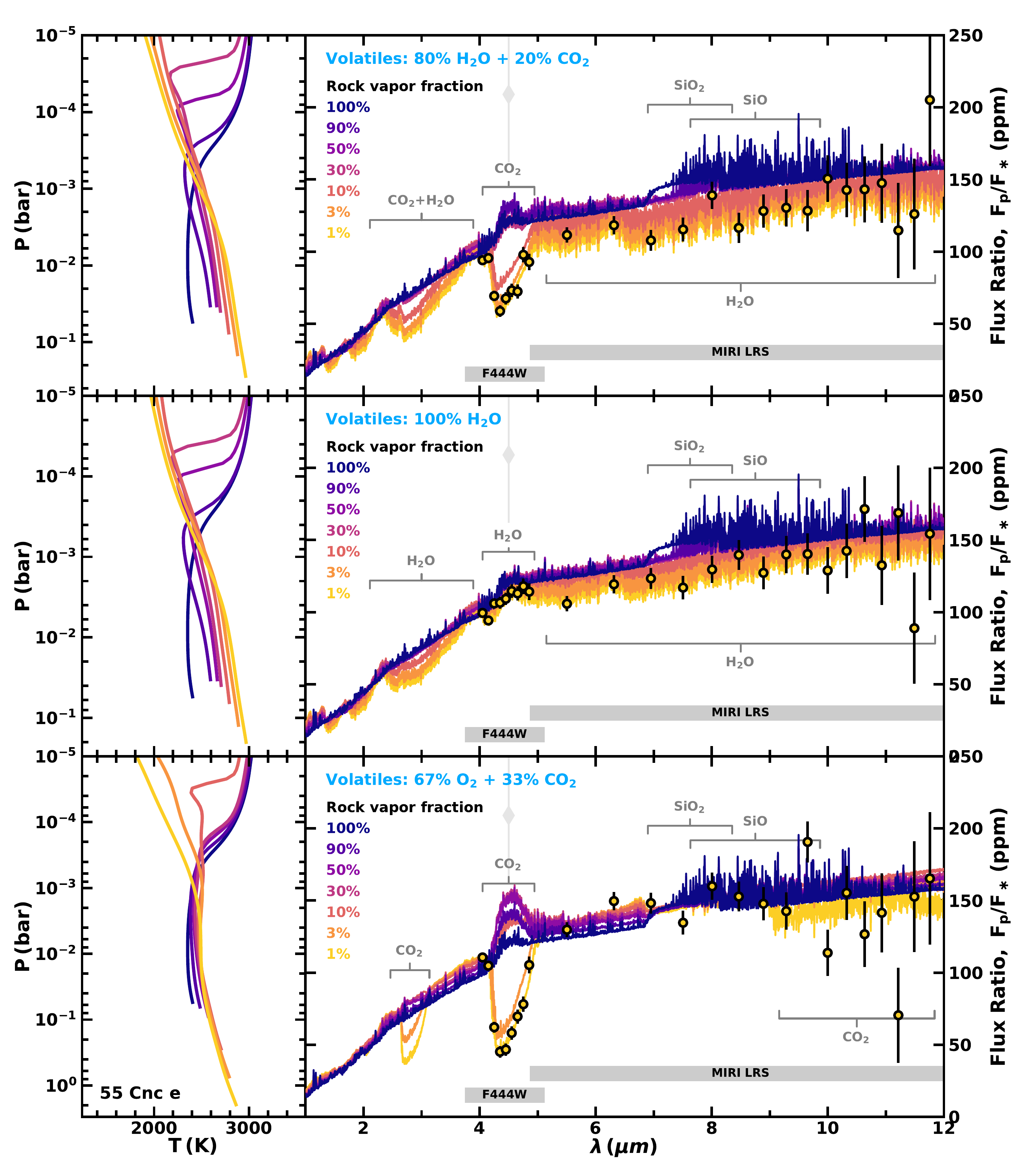}
    \caption{Similar to Figure \ref{fig:HD86226c_results} but for 55~Cnc~e. Simulated JWST observations are shown for the most volatile-rich cases assuming 5 NIRCam~F444W eclipses and one MIRI~LRS eclipse. The light grey diamond and error bars show the existing Spitzer observation using the latest re-analysis from \citet{Mercier2022}. Key spectral features are labeled in grey.}
    \label{fig:55Cnce_results}
\end{figure*}

We assess the observability of rock vapor-rich to volatile-rich atmospheres for the three low-density lava world case studies described in Section \ref{sec:target_selection}: HD~86226~c, HD~3167~b and 55~Cnc~e. For each target, we consider mixed rock vapor/volatile atmospheric compositions, including three different compositions for the volatile component as described in Section \ref{sec:opacity}: 80\% H$_2$O $+$ 20\% CO$_2$, 100\% H$_2$O and 67\% O$_2$ $+$ 33\% CO$_2$. We additionally consider rock vapor fractions from 100--1\% (i.e., volatile fractions of 0--99\%) for HD~3167~b and 55~Cnc~e. For HD~86226~c, we consider rock vapor fractions from 90--1\% as 2D and surface effects become important for the 100\% rock vapor case, given the lower dayside temperature of this planet (see Section \ref{sec:discussion}).

We calculate self-consistent temperature profiles and thermal emission spectra for each atmospheric composition; these are shown in Figures \ref{fig:HD86226c_results}, \ref{fig:HD3167b_results} and \ref{fig:55Cnce_results} for HD~86226~c, HD~3167~b and 55~Cnc~e, respectively. In what follows, we discuss the effects of atmospheric composition (Section \ref{sec:results:composition}) and dayside temperature (Section \ref{sec:results:temperature}) on the temperature structures and observable spectra of these low-density lava worlds. In Section \ref{sec:results:observability}, we discuss the observability of spectral features due to volatile species, and assess the JWST observations required to constrain their presence.

\subsection{Effects of atmospheric composition}
\label{sec:results:composition}

We investigate the effects of atmospheric composition along two axes: the ratio of rock vapor to volatile species in the atmosphere, and the composition of the volatile component. In Figures \ref{fig:HD86226c_results}, \ref{fig:HD3167b_results} and \ref{fig:55Cnce_results}, different rock vapor/volatile ratios are shown by different colors, while the top, middle and bottom panels correspond to models with volatile components consisting of 80\% H$_2$O $+$ 20\% CO$_2$, 100\% H$_2$O and 67\% O$_2$ $+$ 33\% CO$_2$, respectively.

Across the three volatile compositions we consider, we find that rock vapor-rich atmospheric compositions typically have thermally-inverted atmospheres, while volatile-rich atmospheres have temperatures decreasing with altitude. This is a result of the optical/UV and infrared opacity contributions of the rock vapor and volatile components. A thermal inversion occurs when the optical/UV opacity dominates over the infrared opacity in the upper atmosphere \citep{Hubeny2003}. In particular, the strong optical/UV opacity of rock vapor species such as SiO, TiO, Na and K (Figure \ref{fig:opac}) is known to produce strong thermal inversions in pure rock vapor atmospheres \citep{Ito2015,Zilinskas2022}. However, volatile species such as H$_2$O and CO$_2$ contribute significant infrared opacity which can result in a decreasing temperature profile if these species are sufficiently abundant. 

The abundance of volatiles required to transition from a thermally inverted to a non-inverted temperature profile depends on the composition of the volatile component and the dayside temperature, as these factors control the atmospheric chemistry and the infrared vs optical/UV opacity in the atmosphere. For example, for HD~3167~b we find that when the volatile content of the atmosphere consists of 80\% H$_2$O $+$ 20\% CO$_2$ or 100\% H$_2$O, a volatile mixing ratio as low as 10\% (i.e., a rock vapor fraction of 90\%) results in a temperature profile which is non-inverted in the infrared photosphere (though a thermal inversion is still present at lower pressures, this does not affect the infrared emission spectrum). These temperature profiles lead to absorption features due to H$_2$O and/or CO$_2$ in the infrared thermal emission spectrum. As the volatile fraction increases beyond 10\%, the temperature profile decreases slightly more steeply with altitude, the base of the 0.2--50~$\mu$m photosphere extends to deeper pressures, and the absorption features in the spectrum become gradually more pronounced.

When the volatile content of the atmosphere consists of 67\% O$_2$ $+$ 33\% CO$_2$, a higher volatile fraction is required for a non-inverted temperature profile compared to the cases with H$_2$O. This is because H$_2$O provides significant opacity continuously across the infrared, while O$_2$ has very low opacity in comparison. As a result, for HD~3167~b, only volatile fractions $>70$\% (i.e., rock vapor fractions $<$30\%) lead to non-inverted temperature profiles and absorption features in the thermal emission spectrum. For compositions with $\sim$10--70\% volatiles, the thermally-inverted temperature profile leads to emission features due to CO$_2$, most clearly visible at $\sim$4.5~$\mu$m. SiO$_2$ and SiO emission features are also visible at $\sim$7--10~$\mu$m for these thermally inverted atmospheres. For rock vapor fractions $<$30\%, the non-inverted temperature profile leads to CO$_2$, SiO$_2$ and SiO absorption features. 

The presence of SiO$_2$ and SiO features at $\sim$7--10~$\mu$m is sensitive to the presence of H$_2$O in the atmosphere \citep[see also][]{Zilinskas2023}. For the volatile compositions including H$_2$O (top two panels of figures \ref{fig:HD86226c_results}--\ref{fig:55Cnce_results}), the H$_2$O opacity obscures the $\sim$7--10~$\mu$m SiO$_2$ and SiO features, even for high rock vapor fractions. Meanwhile, in the 67\% O$_2$ $+$ 33\% CO$_2$ case (bottom panels of figures \ref{fig:HD86226c_results}--\ref{fig:55Cnce_results}), the silicate features dominate in the $\sim$7--10~$\mu$m range and result in emission or absorption features even for rock vapor fractions of 1\%. Indeed, Figure \ref{fig:opac} shows that H$_2$O is the main opacity source overlapping with the $\sim$7--10~$\mu$m SiO$_2$ and SiO features. The presence of spectral features due to these silicates can therefore act as an indicator that H$_2$O is not significantly present in the atmosphere. As a result, searches for silicates could be used to place constraints on the atmospheric loss of hydrogen. For example, a detection of CO$_2$ combined with detections of SiO and/or SiO$_2$ would imply a volatile-rich interior composition with no H$_2$O, suggesting efficient hydrogen escape.

\subsection{Effects of dayside temperature}
\label{sec:results:temperature}

\begin{figure}
    \centering
    \includegraphics[width=0.5\textwidth]{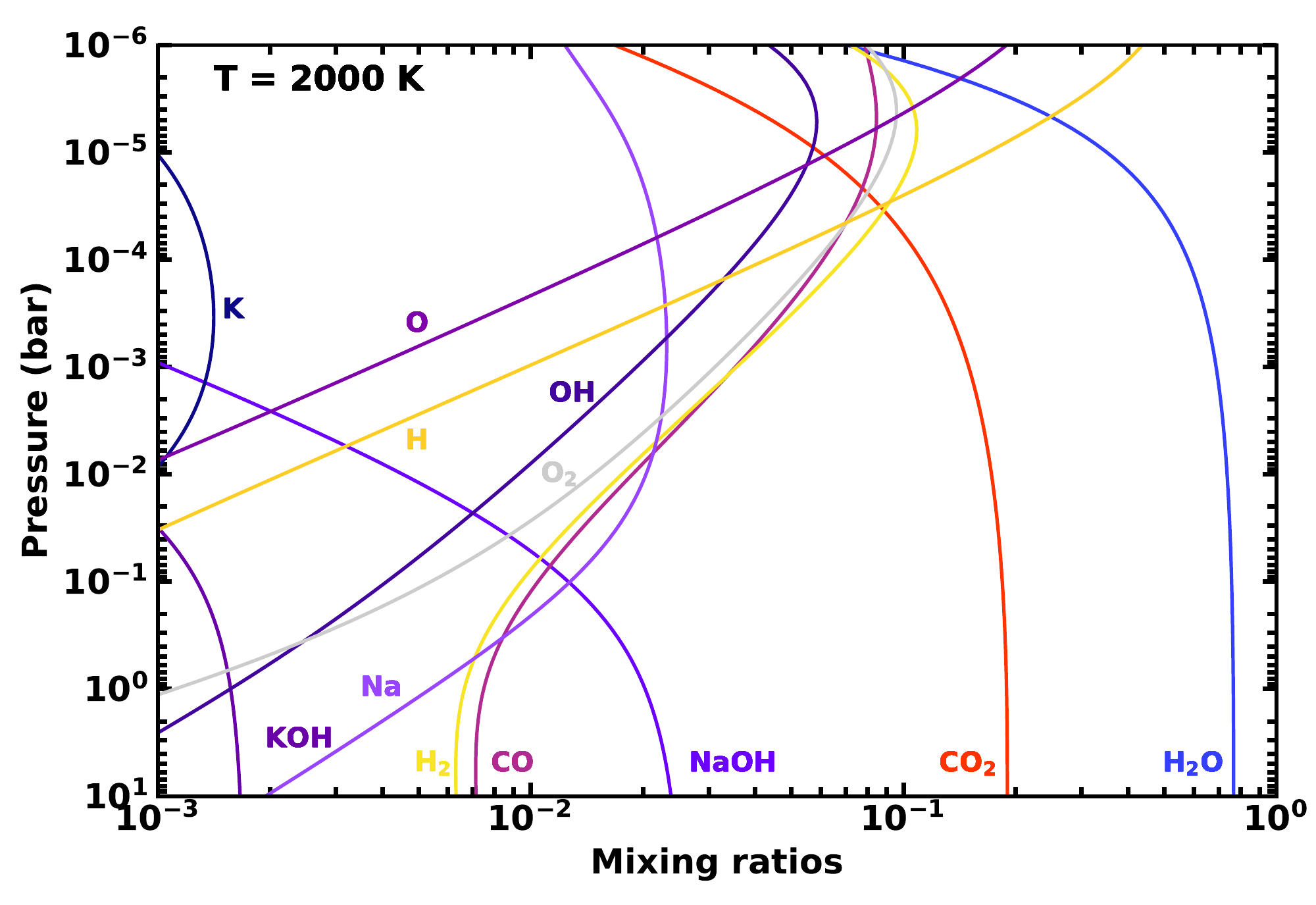}
    \includegraphics[width=0.5\textwidth]{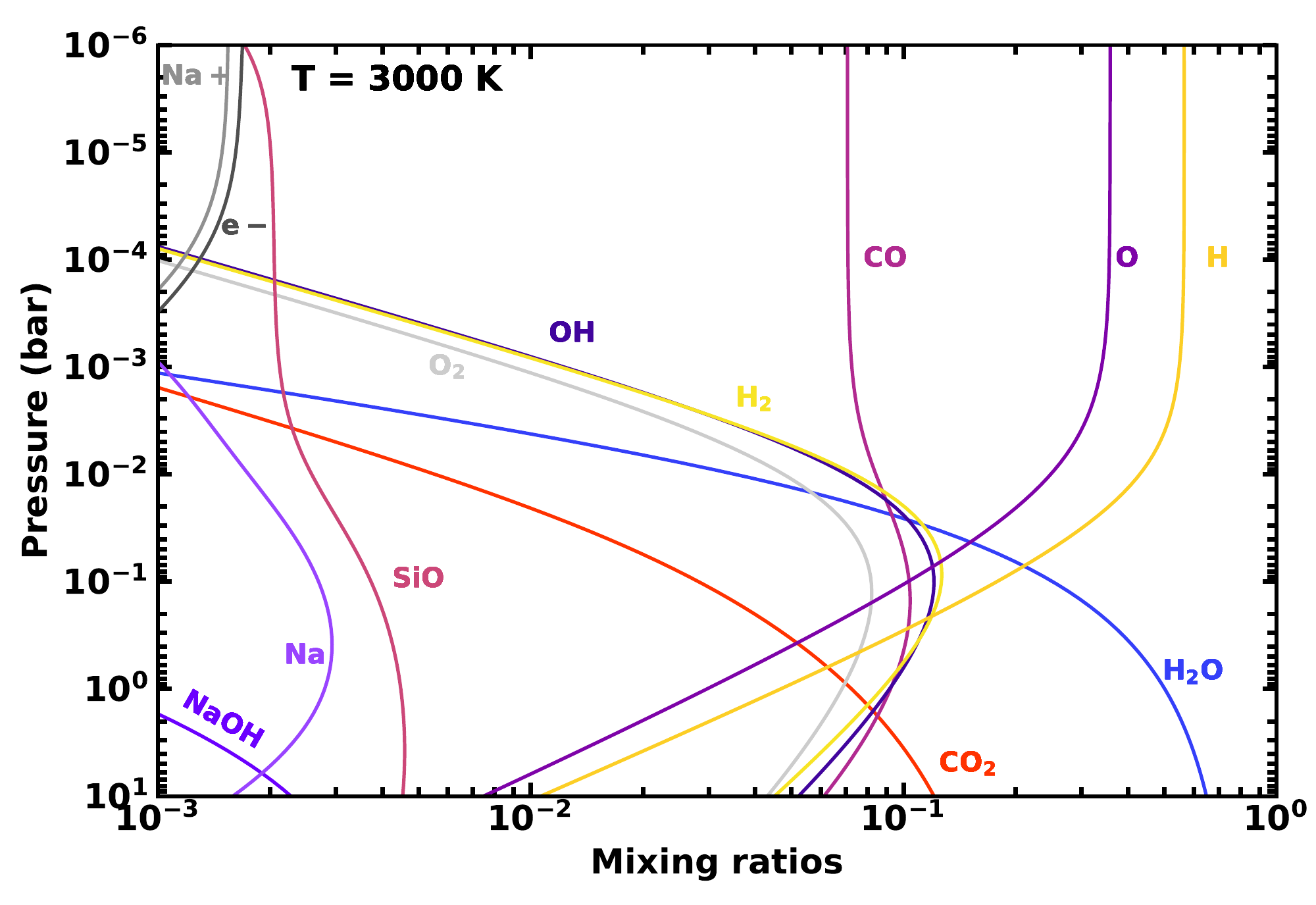}
    \caption{Equilibrium chemical abundances as a function of pressure corresponding to isothermal temperature profiles of 2000~K (top panel) and 3000~K (bottom panel). The elemental abundances correspond to a 1\% rock vapor fraction (99\% volatiles) and a volatile composition consisting of 80\% H$_2$O $+$ 20\% CO$_2$, as described in Section \ref{sec:opacity}.}
    \label{fig:abundances}
\end{figure}

The three case studies we consider span a range of sub-stellar temperatures: 1854~K (HD~86226~c), 2513~K (HD~3167~b) and 2773~K (55~Cnc~e). Increasing temperatures impact the chemistry of the atmosphere, for example through the dissociation of species such as H$_2$O and by changing the outgassed rock vapor composition. In turn, this affects the resulting atmospheric opacity, temperature profile and emission spectrum.

Figure \ref{fig:abundances} shows the impact of temperature and pressure on the equilibrium abundances of the dominant chemical species in a 99\% volatile atmosphere. Abundances are shown for an 80\% H$_2$O $+$ 20\% CO$_2$ volatile composition. For simplicity, these abundances are calculated as a function of pressure for isothermal temperature profiles at 2000~K and 3000~K, respectively. The abundances of H$_2$O and CO$_2$ differ significantly between these two temperature regimes: at 2000~K, H$_2$O and CO$_2$ are the most abundant species down to pressures of $\sim 10^{-4}-10^{-5}$~bar. However, at 3000~K, H, O and CO become the dominant species for pressures below $\sim 1-10^{-2}$~bar, as H$_2$O is dissociated and the favored carbon species becomes CO over CO$_2$. The abundances of rock vapor species such as SiO and Na also differ significantly between the two temperature regimes. In particular, for temperatures above $\sim$2700~K, SiO becomes the most abundant outgassed species from a bulk silicate earth rock composition, whereas Na is the most dominant for lower temperatures \citep[e.g.][]{Miguel2011,vanBuchem2022,Wolf2022}. 

These temperature-dependent equilibrium chemical abundances have significant effects on the temperature structures and thermal emission spectra of low-density lava worlds. Notably, the depletion of H$_2$O and CO$_2$ at higher temperatures significantly reduces the infrared opacity of the atmosphere, leading to more inverted temperature profiles. This means that the transition from an inverted to a non-inverted temperature profile occurs at higher volatile fractions for hotter planets. For example, for HD~86226~c in the case of an 80\% H$_2$O $+$ 20\% CO$_2$ volatile composition, a 10\% volatile mixing ratio (90\% rock vapor, dark blue line in top panel of Figure \ref{fig:HD86226c_results}) is sufficient to suppress a thermal inversion and leads to H$_2$O and CO$_2$ absorption features. Conversely, volatile fractions of 50\% and 90\% are required for HD~3167~b and 55~Cnc~e, respectively, to suppress a thermal inversion. The detection of absorption features in the atmospheric spectrum would therefore imply different volatile fractions for each of these planets, depending on their dayside temperatures.

\subsection{Atmospheric retrievals and observability}
\label{sec:results:observability}

\begin{figure*}
    \centering
    \includegraphics[width=\textwidth]{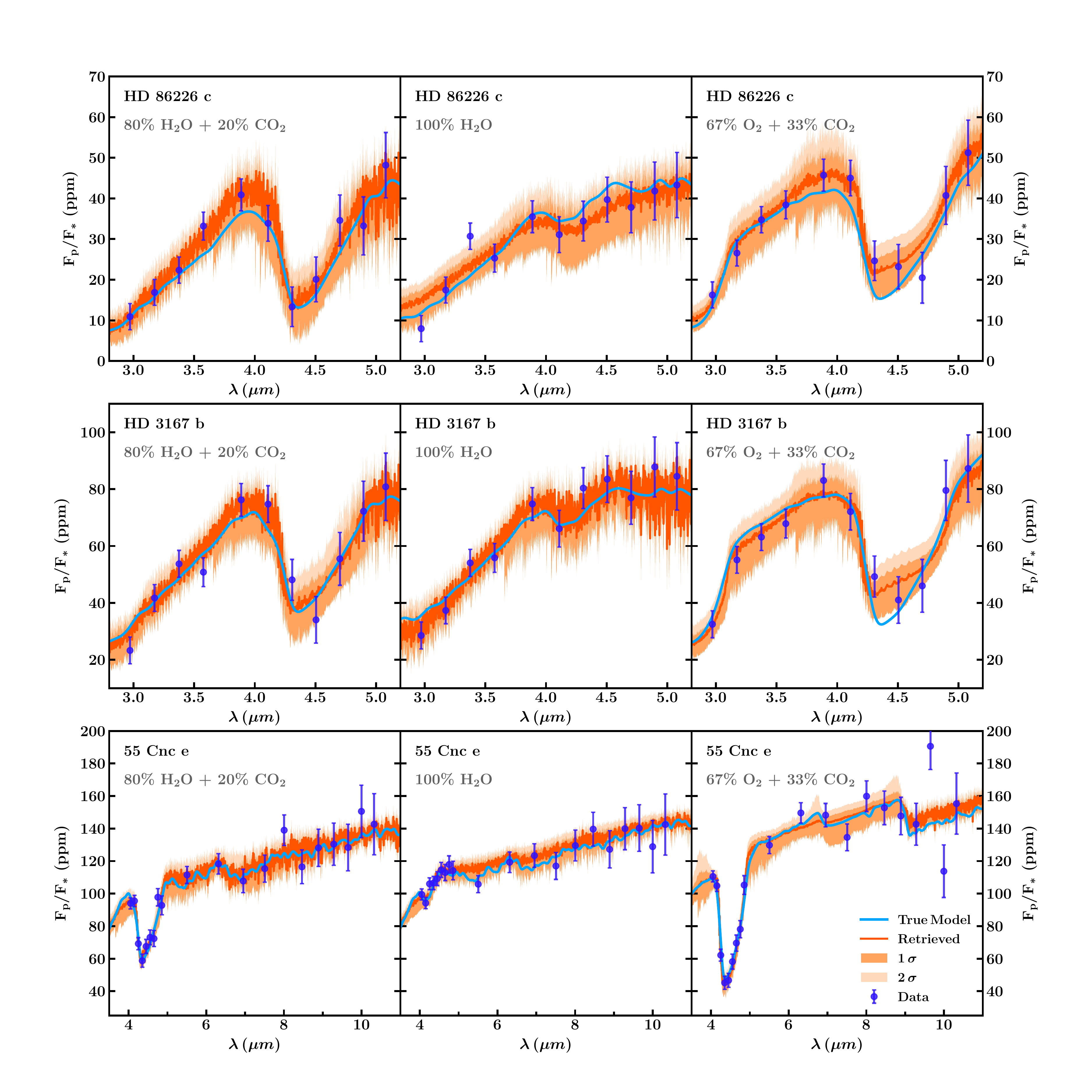}
    \caption{Retrieved thermal emission spectra for the simulated JWST observations shown in Figures \ref{fig:HD86226c_results}, \ref{fig:HD3167b_results} and \ref{fig:55Cnce_results} for HD~86226~c (top row), HD~3167~b (middle row) and 55~Cnc~e (bottom row), respectively. The simulated observations correspond to an atmospheric composition with 1\% rock vapor and 99\% volatiles, where the volatile composition is either 80\% H$_2$O $+$ 20\% CO$_2$ (left column), 100\% H$_2$O (middle column) or 67\% O$_2$ $+$ 33\% CO$_2$ (right column). Purple points and error bars show the simulated observations, while solid blue lines show the `true' underlying model. The medium retrieved spectra and 1$\sigma$/2$\sigma$ contours are shown as orange lines and dark/light orange shading, respectively. Note that the bottom row spans a different wavelength range to the top two rows.}
    \label{fig:ret_spec}
\end{figure*}

The thermal emission spectra in Figures \ref{fig:HD86226c_results}--\ref{fig:55Cnce_results} show a number of spectral features due to H$_2$O, CO$_2$, SiO and SiO$_2$. In particular, H$_2$O has features visible at 2--4~$\mu$m and 5--12~$\mu$m, while CO$_2$ has strong features at $\sim$3~$\mu$m and $\sim$4.5~$\mu$m, and a broader opacity feature at $\sim$9--12~$\mu$m. Features due to SiO$_2$ and SiO are also visible at $\sim$7--8~$\mu$m and $\sim$8--10~$\mu$m, respectively. However, the strength and observability of these features varies with atmospheric composition and dayside temperature.

We find that the strongest and most easily observable spectral features for these low-density lava worlds are absorption features due to CO$_2$ and H$_2$O in the case of a volatile-rich atmosphere. The volatile mixing ratio required to form these features depends on the atmospheric composition and dayside temperature, as discussed in sections \ref{sec:results:composition} and \ref{sec:results:temperature}. However, beyond this threshold, higher volatile mixing ratios result in stronger H$_2$O and/or CO$_2$ absorption features. Besides volatile species, spectral features due to SiO and SiO$_2$ can be seen in the $\sim$7--10~$\mu$m range, though these features can be washed out by H$_2$O opacity, as discussed in Section \ref{sec:results:composition}. In what follows, we focus on the observability of volatile spectral features, namely H$_2$O and CO$_2$, whose detection in a low-density lava world atmosphere could indicate a volatile-rich interior composition.

In order to assess the observability of these molecular features, we simulate JWST observations for each of the three case studies modeled. We simulate these observations across the three volatile compositions considered above (80\% H$_2$O $+$ 20\% CO$_2$, 100\% H$_2$O and 67\% O$_2$ $+$ 33\% CO$_2$) assuming a 1\% rock vapor fraction (i.e., a 99\% volatile fraction), as this results in the strongest H$_2$O and CO$_2$ absorption features. For HD~86226~c and HD~3167~b, we model NIRSpec~G395H and MIRI~LRS secondary eclipse spectra assuming five eclipses for each observing mode. For 55~Cnc~e, we simulate secondary eclipse spectra assuming five eclipses with NIRCam~F444W and one eclipse with MIRI~LRS, as these observations will be acquired in JWST's Cycle 1. These simulated observations are shown in Figures \ref{fig:HD86226c_results}--\ref{fig:55Cnce_results}. We note that, while the MIRI~LRS observation of 55~Cnc~e is expected to saturate the detector at shorter wavelengths and therefore reduce the usable wavelength range, we have not accounted for this in our simulated observations and therefore present the full 5--12~$\mu$m wavelength range. HD~86226~c, HD~3167~b and 55~Cnc~e have very short transit durations of 3.1, 1.6 and 1.6 hours, respectively \citep{Demory2011,Teske2020,Bourrier2022}. Observing $\sim$5 eclipses for one of these targets is therefore feasible within a small to medium JWST proposal, making them excellent targets for investigating super-Earth interior compositions.

We perform atmospheric retrievals on the simulated observations to assess the confidence with which H$_2$O and/or CO$_2$ could be detected for each case study and each volatile composition. For HD~86226~c and HD~3167~b, we use only the five NIRSpec~G395H simulated observations for the retrievals, as these are sufficient to constrain H$_2$O and CO$_2$ without the need for a further five eclipses with MIRI~LRS. In particular, we find that for HD~3167~b, H$_2$O and CO$_2$ are detectable (where present) with $>$3$\sigma$ confidence across the three volatile compositions. For HD~86226~c, $>$3$\sigma$ detections of H$_2$O and/or CO$_2$ are possible for the 80\% H$_2$O $+$ 20\% CO$_2$ and 67\% O$_2$ $+$ 33\% CO$_2$ volatile compositions, while a tentative $\sim$2$\sigma$ detection of H$_2$O is possible for the 100\%~H$_2$O volatile composition. 

\begin{figure*}
    \centering
    \includegraphics[width=\textwidth]{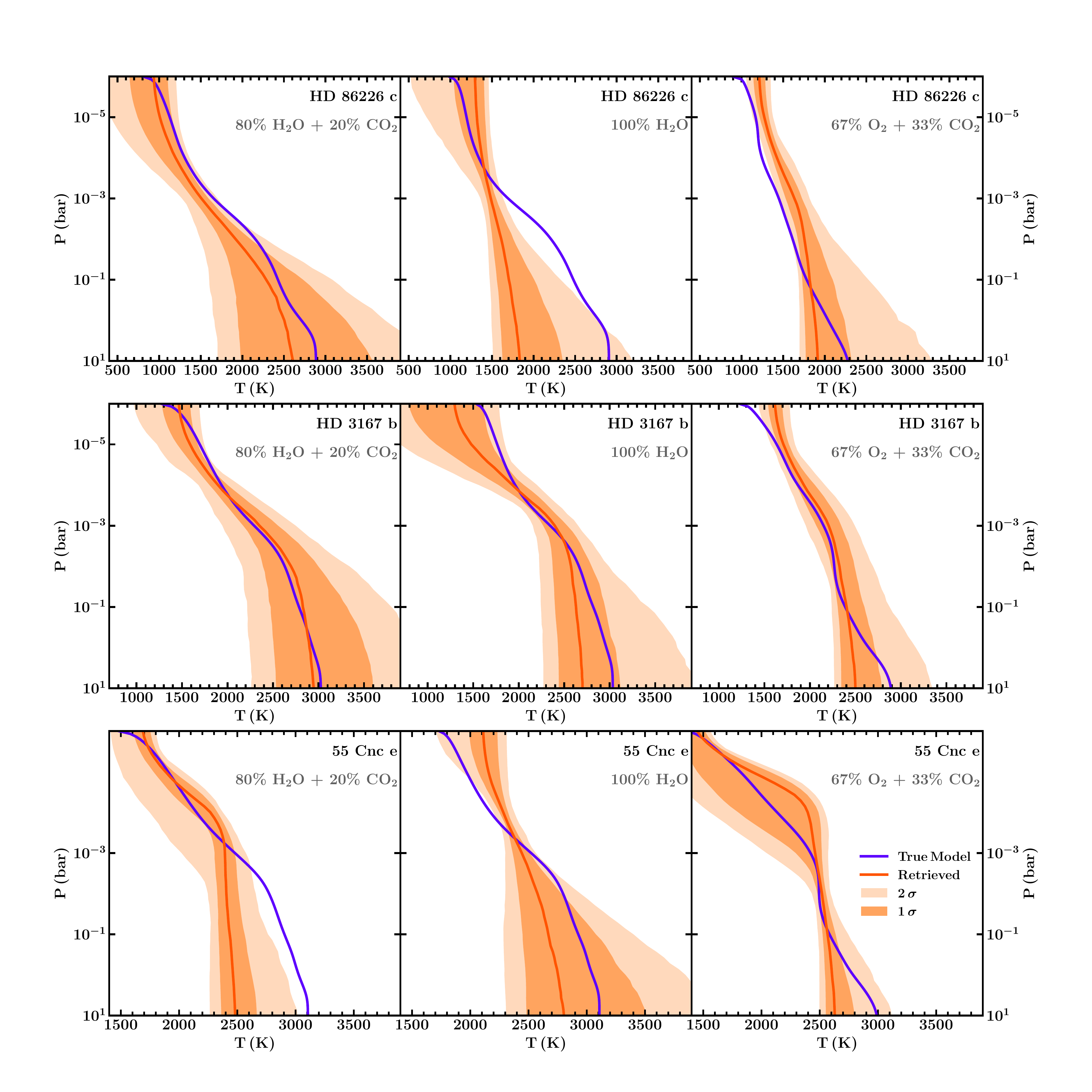}
    \caption{Retrieved atmospheric temperature profiles for HD~86226~c (top row), HD~3167~b (middle row) and 55~Cnc~e (bottom row), respectively, given the simulated JWST observations shown in Figures \ref{fig:HD86226c_results}, \ref{fig:HD3167b_results} and \ref{fig:55Cnce_results}. The simulated observations correspond to an atmospheric composition with 1\% rock vapor and 99\% volatiles, where the volatile composition is either 80\% H$_2$O $+$ 20\% CO$_2$ (left column), 100\% H$_2$O (middle column) or 67\% O$_2$ $+$ 33\% CO$_2$ (right column). Solid purple lines show the `true' model temperature profile. The medium retrieved temperature profiles and 1$\sigma$ and 2$\sigma$ contours are shown as orange lines and dark/light orange shading, respectively.}
    \label{fig:ret_pt}
\end{figure*}

\begin{figure*}
    \centering
    \includegraphics[width=0.75\textwidth]{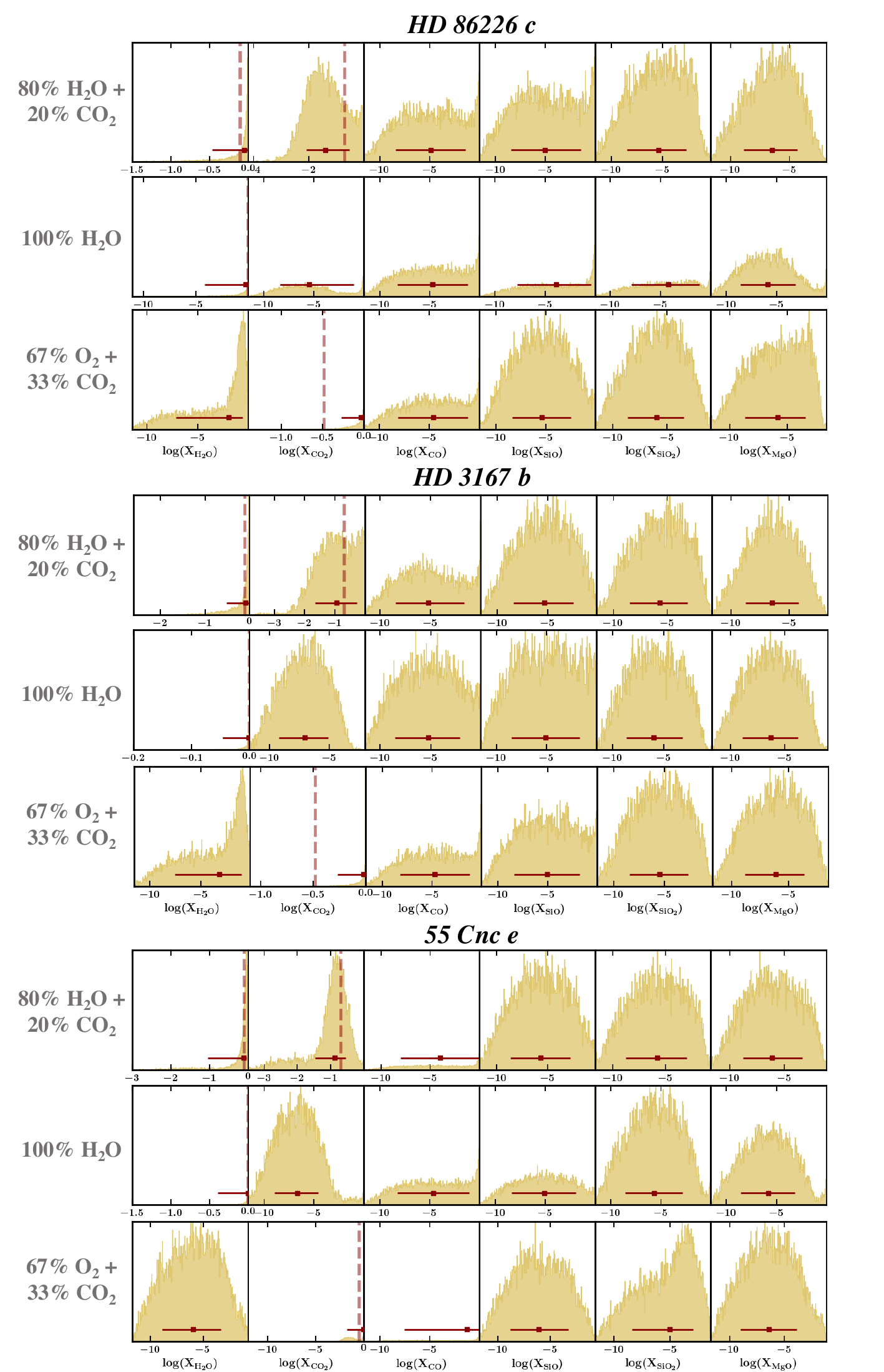}
    \caption{Retrieved posterior probability distributions for the molecular volume mixing ratios. The posterior distributions correspond to the retrieval cases in Figures \ref{fig:ret_spec} and \ref{fig:ret_pt} for HD~86226~c (top three rows), HD~3167~b (middle three rows) and 55~Cnc~e (bottom three rows).  Red squares and error bars show the median retrieved abundances and 68\% confidence intervals, respectively. Dashed vertical lines show the nominal H$_2$O and CO$_2$ abundances, without accounting for thermal dissociation effects.}
    \label{fig:ret_post}
\end{figure*}

Figures \ref{fig:ret_spec}, \ref{fig:ret_pt} and \ref{fig:ret_post} show the retrieved spectra, temperature profiles and abundance posteriors, respectively. We find that the retrieval is able to accurately retrieve the input temperature profile and spectrum. While the abundance constraints we obtain are generally quite broad, the retrieval is able to identify H$_2$O-dominated compositions and the presence of CO$_2$, and is generally consistent with the input abundances within $\sim$2$\sigma$. When the atmospheric composition is dominated by O$_2$ and CO$_2$, the retrieval typically overestimates the abundance of CO$_2$ as this is the only spectrally active species in the wavelength range probed. Nevertheless, the retrieval is able to confidently detect the presence of CO$_2$ in the atmosphere in this scenario. We additionally include CO, SiO, SiO$_2$ and MgO in our retrievals, but these species are not constrained due to their lower abundances in the input spectra and/or a lack of spectral features in the wavelength range of the simulated observations. 

In the case of 55~Cnc~e, we find that CO$_2$ can be detected very confidently (at $>$8$\sigma$ confidence), but detections of H$_2$O are more tentative (2.5--3$\sigma$ confidence). The lower confidence in these H$_2$O detections is partly a result of the wavelength range of the 55~Cnc~e simulated observations; due to saturation issues, the 2--4~$\mu$m H$_2$O spectral feature is not accessible for this target with JWST. While MIRI~LRS is sensitive to H$_2$O opacity, this broad feature is more challenging to constrain given the observational uncertainties. Furthermore, as discussed in Section \ref{sec:results:temperature}, H$_2$O is expected to be somewhat depleted due to thermal dissociation in the atmosphere of 55~Cnc~e due to its extremely high dayside temperature.

These results represent the minimum observing time required to detect H$_2$O and/or CO$_2$ for the three targets, as the 1\% rock vapor (99\% volatiles) models have the strongest absorption features due to these species. However, atmospheres with lower volatile fractions could also have detectable H$_2$O and/or CO$_2$ spectral features. For example, in the case of HD~86226~c, models with $\gtrsim$50\% H$_2$O and/or CO$_2$ (top two panels of Figure \ref{fig:HD86226c_results}) have H$_2$O and CO$_2$ features similar in size to the 99\% volatile case. When the volatiles consist of O$_2$ + CO$_2$, large CO$_2$ absorption features are visible for $\gtrsim$90\% atmospheric volatile fractions, again similar in size to the 99\% volatile case. Furthermore, in the case of a 10\% volatile fraction, a significant CO$_2$ emission feature is present at $\sim$4.5~$\mu$m with a signal-to-noise ratio (S/N) of $\gtrsim 3$ assuming 5 NIRSpec~G395H eclipses. The 5 NIRSpec~G395H eclipses discussed above would therefore be sufficient to characterize a range of atmospheric compositions and volatile fractions for HD~86226~c. In the case of HD~3167~b, atmospheric volatile fractions $\gtrsim$90\% have similar H$_2$O and/or CO$_2$ feature sizes as the 99\% volatile cases across all three volatile compositions. The 5 NIRSpec~G395H eclipses proposed above would therefore be sensitive to volatile fractions $\gtrsim$90\%. For 55~Cnc~e, the excellent S/N of the observations should allow H$_2$O and/or CO$_2$ to be detected across an even wider range of volatile fractions.

Spectral features due to SiO and/or SiO$_2$ may also be detectable in the mid-infrared using MIRI~LRS. For example, in the case of HD~86226~c with $\gtrsim$70\% O$_2$ + CO$_2$ (bottom panel of Figure \ref{fig:HD86226c_results}), a strong blended SiO$_2$/SiO absorption feature is present at $\sim$7-9~$\mu$m with S/N$\gtrsim$3, assuming 5 MIRI~LRS eclipses. In the case of a pure rock vapor atmospheric composition, the lower S/N and lower resolution of MIRI~LRS makes it more challenging to detect the SiO$_2$/SiO features. However, in this volatile-free scenario, surface effects and atmospheric dynamics can become important \citep[e.g.,][]{Nguyen2022}. The observability of these species may therefore be greater than predicted here. We discuss this further in Section \ref{sec:discussion}.

\section{Discussion}
\label{sec:discussion}

Our results show that volatile-rich atmospheres around low-density lava worlds can be identified using their infrared emission spectra, providing a method to test for volatile-rich interior compositions. However, the threshold abundance of volatiles needed for H$_2$O and/or CO$_2$ spectral features to be visible depends on the dayside temperature and the composition of the volatile component. The relative abundances of rock vapor and volatiles in these atmospheres depends on a range of factors, including how volatiles are stored in the interior, the relative solubilities of different volatile species in a magma ocean, as well as atmospheric escape rates for different chemical species. While these factors and their relative influences on the atmospheric composition remain largely unconstrained, we discuss their potential effects and the future work which is needed to understand them. 

Volatiles can be stored in different parts of the planetary interior, including the upper mantle, the deep mantle and the iron core \citep{Li_2019_bookchapter,Dorn2021,Schlichting2022,Kovacevic2022,Vazan2022}. In the mantle, lava worlds are expected to have deep dayside magma oceans, reaching as deep the core-mantle boundary in some cases \citep{Boukare2022}. The availability of volatiles to evaporate/degas at the surface therefore depends on the efficiency of mixing in the magma ocean; for example, \citet{Boukare2022} find that vigorous convection occurs if the core temperature is sufficiently high. Furthermore, buoyancy effects may cause volatile-rich material to rise to the surface, though this depends on its material properties as a function of pressure \citep[e.g.,][]{Karki2021}. If the bulk densities of low-density lava worlds are a result of volatile-rich mantles, it may therefore be reasonable to expect volatiles at the planetary surface. Furthermore, if volatiles are stored in the deep mantle, they are likely to contribute less to the low bulk density of the planet given the high pressures in this region, though this is dependent on high-pressure material properties. The low bulk densities measured for several lava worlds may therefore require the presence of volatiles in the upper mantle and/or the iron core.

For some low-density lava worlds, a volatile-rich or absent core may explain the bulk density without the need for a volatile-rich mantle. The presence of light elements in the iron core can significantly reduce the bulk density of a super-Earth, with hydrogen resulting in the largest density deficits \citep{Li_2019_bookchapter,Schlichting2022}. Indeed, \citet{Schlichting2022} show that if Fe metal is able to participate chemically as the planetary interior equilibrates, the iron core can consist of several tens of percent of H and O. The resulting mass-radius relations are consistent with low-density lava worlds such as HD~3167~b and 55~Cnc~e, though some planets with even lower bulk densities (e.g., HD~86226~c) cannot be explained with this mechanism. Similarly, the absence of an iron core can explain the low bulk densities of HD~3167~b and 55~Cnc~e, but this is insufficient to explain a planet less dense than a 100\% silicate composition (e.g., HD~86226~c), which requires additional volatiles. For planets such as HD~3167~b and 55~Cnc~e, the detection of a purely rocky atmosphere may therefore indicate a volatile-rich or absent core, with implications for their formation and evolution mechanisms.

If volatiles are indeed present in the mantles of low-density lava worlds, these can be dissolved in a `wet' magma ocean \citep{Dorn2021}, or become miscible with rock to form a mixed-composition layer \citep{Kovacevic2022,Vazan2022}. In either scenario, both volatiles and rock species can evaporate into the atmosphere due to the high dayside temperature. However, the relative abundances of these two components may vary. For example, for a miscible water/rock mantle with a surface temperature $\gtrsim$2000~K, miscibility can extend to pressures as low as 100~bar \citep{Vazan2022}; at lower temperatures and pressures, the two components can de-mix, resulting in a thick steam-dominated atmosphere. Conversely, in the case of a wet magma ocean, the atmospheric abundances of species such as H$_2$O and CO$_2$ depend on their solubilities in the magma ocean, which vary with mantle composition and surface pressure and temperature \citep[e.g.,][]{Lichtenberg2021a,Bower2022}. For example, given the higher solubility of H$_2$O in basaltic magma oceans compared to CO$_2$, CO$_2$ may be preferentially partitioned into the atmosphere \citep[e.g.,][]{Bower2022}.

Atmospheric escape mechanisms are also expected to affect the relative abundances of rock vapor and volatiles in these atmospheres. While the evaporation of surface material can replenish material lost from the atmosphere, differential mass loss rates for different chemical species may impact the overall atmospheric composition. The high irradiation levels received by lava worlds are conducive to atmospheric loss, but several works have shown that radiative line cooling can significantly mitigate this loss. In particular, \citet{Ito2021} find that species such as Na, Mg, Mg$^+$, Si$^{2+}$, Na$^{3+}$, and Si$^{3+}$ are able to efficiently cool the atmospheres of hot rocky exoplanets with rock vapor atmospheres, leading to reduced atmospheric mass loss rates. Similarly, radiative cooling due to H$_2$O, CO$_2$ and their dissociation products (e.g., OH, H$_3^+$ and CO) can also limit atmospheric mass loss \citep{Tian2009,Yoshida2022}. Future work will be required to understand how competing mass loss and radiative cooling effects influence the compositions of mixed volatile/rock vapor atmospheres.

The observability of low-density lava world atmospheres is influenced by several factors which we have not considered in this work. For example, non-local thermodynamic equilibrium (non-LTE) effects are known to influence the temperature profiles and radiative transfer in the low-density regions of planetary atmospheres \citep{Young2020}. Given the low photospheric pressures of rock vapor-rich atmospheres, non-LTE effects may have a non-negligible influence on their temperature profiles and spectra. However, the more volatile-rich compositions we consider have deeper photospheres (as deep as a few bar in some cases), where such effects may be less important. While non-LTE calculations are beyond the scope of this work, future investigation of such models will be valuable to understand the observable spectra of low-density lava worlds.

2D effects can also influence the atmospheric structures and observability of lava world atmospheres. In the case of a pure rock vapor atmosphere, the atmospheric pressure decreases away from the sub-stellar point due to the decreasing surface temperature. This can impact the overall dayside emission spectrum, potentially deviating from hemispherically-averaged 1D models such as those shown here (see, e.g., \citealt{Nguyen2020,Nguyen2022}). Furthermore, regions where the atmosphere is optically thin (e.g., away from the sub-stellar point and/or for dayside temperatures $\lesssim$2000~K) have strong temperature differences between the surface and the base of the atmosphere \citep{Nguyen2022,Zieba2022,Zilinskas2022}. These temperature jumps can lead to strong spectral features, even if the overlying atmosphere is relatively isothermal. The pure rock vapor cases shown in figures \ref{fig:HD3167b_results} and \ref{fig:55Cnce_results} may therefore underestimate the strength of the SiO and SiO$_2$ spectral features. Given the cooler dayside temperature of HD~86226~c, we do not show a 1D model for the pure rock vapor case as such an atmosphere would be optically thin even at the sub-stellar point. The spectrum would therefore be strongly sensitive to surface and 2D effects which are not considered in our model. For the mixed rock vapor/volatile compositions we consider in this work, the atmospheric pressures are likely to be higher compared to purely rocky lava worlds given the lower evaporation temperatures for volatile species such as H$_2$O and CO$_2$. Therefore, the 2D effects described above are expected to be less important for such cases.

We do not consider clouds in this work as several of the cases we consider (namely, for HD~3167~b and 55~Cnc~e) have dayside temperatures which are too hot to condense silicate clouds. However, we note that for cooler cases (e.g., HD~86226~c) silicate clouds could condense in the upper regions of the atmosphere \citep[e.g.][]{Gao2021}. Furthermore, even in the case of a hotter atmosphere, clouds could condense on the nightside or at the terminator and be transported to the dayside. Such clouds could reduce the amplitude of observed gas species spectral features and make them more difficult to detect. However, cloud signatures in these observations would also place new constraints on cloud formation in extreme conditions. Such signatures include silicate cloud spectral features at $\sim$10~$\mu$m \citep{Pinhas2017,Gao2021}, as well as a cooler dayside spectrum resulting from an increased Bond albedo due to cloud scattering.

\section{Summary and Conclusions}
\label{sec:conclusions}

Ultra-hot super-Earths (or `lava worlds') provide a unique window into planetary interior compositions, as their high dayside temperatures cause the surface to evaporate into the atmosphere. While previous works have focused on the atmospheric compositions of purely rocky lava worlds \citep{Miguel2011,Ito2015,Zilinskas2022} or outgassing from a magma ocean into a primary atmosphere \citep{Zilinskas2023}, several known lava worlds are consistent with having volatile-rich interior compositions. In this work, we therefore explore the atmospheric structures and observability of these `low-density lava worlds', whose atmospheres could contain a mixture of rock vapor and volatile species evaporated from the surface.

We use a self-consistent atmospheric model to calculate the equilibrium temperature profiles, equilibrium thermochemical profiles and thermal emission spectra of low-density lava worlds. We consider a range of rock vapor-rich to volatile-rich compositions, from 100\% rock vapor scenarios to 1\% rock vapor plus 99\% volatile. For the volatile component, we consider three different compositions to represent a range of interior compositions and atmospheric evolution scenarios: (i) a mixed 80\% H$_2$O $+$ 20\% CO$_2$ composition, representing a scenario in which carbon-containing volatiles are present as well as H$_2$O, (ii) 100\% H$_2$O, in the case of a pure `water world', and (iii) a 67\% O$_2$ $+$ 33\% CO$_2$ composition, similar to case (i) but assuming that all the hydrogen is lost due to atmospheric escape.

In order to identify ideal targets for atmospheric observations of low-density lava worlds, we assess the observability of known super-Earths with sub-stellar temperatures $>1700$~K and bulk densities between those expected for a 100\% silicate interior and a 100\% H$_2$O interior (Figures \ref{fig:MR} and \ref{fig:targets}). We use the Emission Spectroscopy Metric (ESM, \citealt{Kempton2018}) to broadly assess the observability of thermal emission from these targets based on their temperatures and stellar properties. We find that the most optimal targets include 55~Cnc~e, pi~Men~c, HD~3167~b, HD~86226~c, TOI-500~b and TOI-561~b. For our more detailed observability analysis using atmospheric models and retrievals, we choose to focus on three case studies which span a range of sub-stellar temperatures: HD~86226~c (1854~K), HD~3167~b (2513~K) and 55~Cnc~e (2773~K). We model the atmospheric temperature profiles and thermal emission spectra for these case studies over the range of atmospheric compositions described above.

We find that the temperature structures and thermal emission spectra of low-density lava world atmospheres are strongly dependent on their relative abundances of rock vapor vs volatile species, as well as the composition of this volatile component. While rock vapor-dominated compositions result in thermal inversions and emission features in the secondary eclipse spectrum, volatile-rich compositions result in non-inverted temperature profiles and absorption features. However, the transition from an inverted to a non-inverted temperature profile occurs at different rock vapor to volatile ratios for different volatile compositions. For example, if the volatile component includes H$_2$O, the strong infrared opacity from this molecule is able to cause non-inverted temperature profiles with comparatively low volatile mixing ratios (e.g., $\sim$10\% volatiles for HD~86226~c). Conversely, if the volatile component consists of O$_2$ and CO$_2$, a larger volatile fraction is needed to result in a non-inverted temperature profile (e.g., 50\% volatiles for HD~86226~c).

The presence of emission vs absorption features in the secondary eclipse spectra of low-density lava worlds is also sensitive to the dayside temperature of the planet. At higher temperatures, species such as H$_2$O and CO$_2$ are thermally dissociated, reducing the infrared opacity of the atmosphere and resulting in more isothermal or inverted temperature profiles. For example, if the volatile component consists of 80\% H$_2$O $+$ 20\% CO$_2$, a $\sim$10\%/50\%/90\% volatile mixing ratio is needed for a non-inverted temperature profile in the atmospheres of HD~86226~c/HD~3167~b/55~Cnc~e, respectively. The detection of H$_2$O and/or CO$_2$ absorption features in the emission spectrum of a low-density lava world would therefore place different limits on the atmospheric volatile content depending on its dayside temperature.

We find that the secondary eclipse spectra of atmospheres with mixed volatile/rock vapor compositions have a number of spectral features in the infrared due to species such as H$_2$O, CO$_2$, SiO and SiO$_2$, some of which may be observable with JWST. The strength and observability of these features is dependent on the overall atmospheric composition and the dayside temperature. While pure rock vapor atmospheres have spectral features due to SiO and SiO$_2$ \citep{Zilinskas2022}, we find that these features (e.g., in the $\sim$7--10~$\mu$m range) are washed out if small amounts of H$_2$O are present due to the strong opacity of H$_2$O across the infrared. Conversely, atmospheres consisting of rock vapor, O$_2$ and CO$_2$ can display emission or absorption features due to SiO/SiO$_2$, depending on the relative abundances of rock vapor and O$_2$/CO$_2$. Across a range of volatile compositions, spectral features due to H$_2$O and/or CO$_2$ are also visible. For the O$_2$+CO$_2$ volatile composition, emission features due to CO$_2$ can be present in the case of rock vapor-rich compositions. For volatile-rich compositions with non-inverted atmospheres, H$_2$O and/or CO$_2$ absorption features are present.

We assess the observability of H$_2$O and CO$_2$ absorption features with JWST for the three case studies we consider. To do this, we simulate JWST observations for each of the three volatile compositions described above, assuming a 99\% volatile mixing ratio as this results in the strongest absorption features. For HD~86226~c and HD~3167~b, we find that five secondary eclipses with NIRSpec G395H are sufficient to detect H$_2$O and/or CO$_2$ with statistical significance ($\gtrsim 3\sigma$) for the 80\% H$_2$O $+$ 20\% CO$_2$ and 67\% O$_2$ $+$ 33\% CO$_2$ volatile compositions. For the 100\% H$_2$O composition, H$_2$O is more tentatively detected at $\gtrsim$2--3$\sigma$. In the case of 55~Cnc~e, we predict that the planned JWST Cycle 1 observations (five NIRCam~F444W eclipses and one MIRI~LRS eclipse) should be sufficient to detect H$_2$O and/or CO$_2$ across these three volatile compositions, assuming a $\geq$99\% volatile mixing fraction. In the case where volatiles are not detected, constraints could still be placed on the atmospheric composition by looking for spectral features due to SiO and SiO$_2$ in the 7--10~$\mu$m range.

Thanks to their extreme conditions, low-density lava worlds represent unique laboratories to probe the interior compositions of super-Earths. In particular, searching for signs of volatile species from their evaporating surfaces could provide new evidence for volatile-rich interiors among the enigmatic super-Earth population.

\begin{acknowledgments}
This AEThER publication is funded in part by the Alfred P. Sloan Foundation under grant G202114194. TL was supported by a research grant from the Branco Weiss Foundation. We thank the anonymous referee for their helpful comments on this manuscript.
\end{acknowledgments}

\appendix

\section{Effect of surface temperature on outgassed chemistry}
\label{sec:appendix:Tsurf}
\begin{figure*}
\centering
    \includegraphics[width=\textwidth]{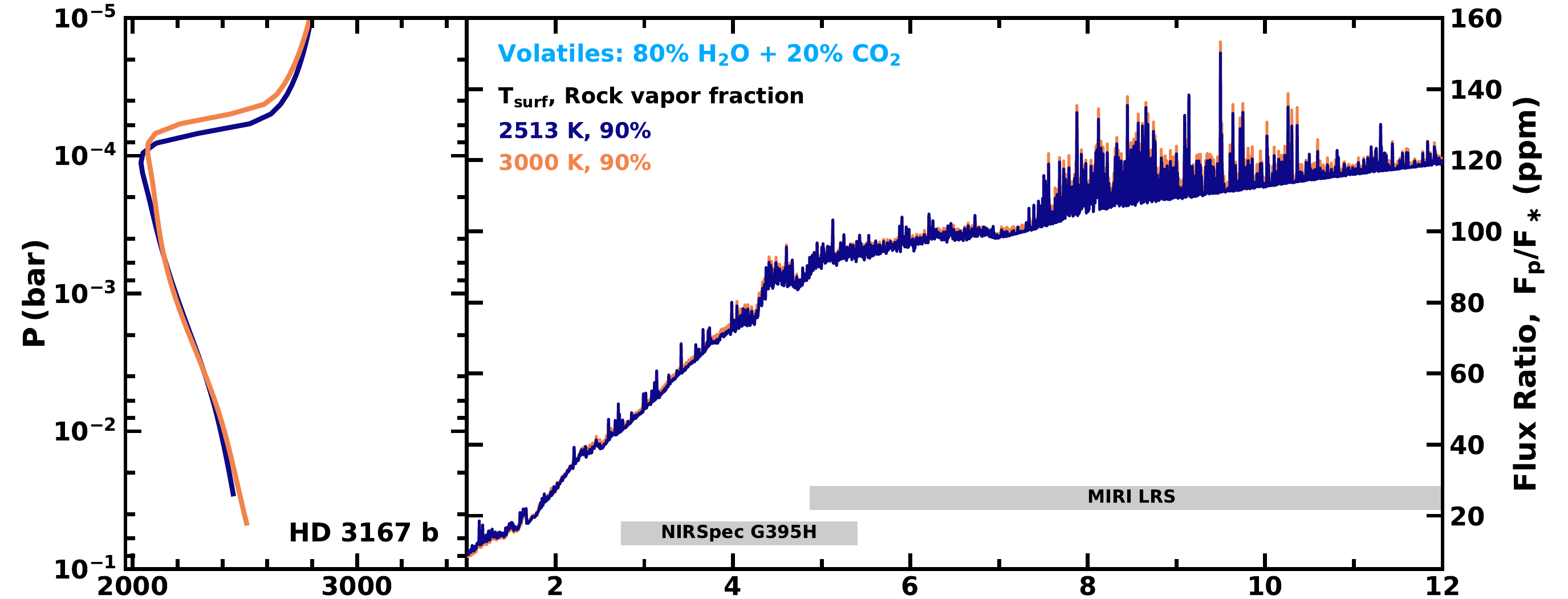}
    \caption{The effect of varying the input surface temperature in the \texttt{VapoRock} calculation on the resulting atmospheric temperature profile and emission spectrum. The example shown is for HD~3167~b assuming a 90\% rock vapor fraction, with the volatile component consisting of 80\% H$_2$O + 20\% CO$_2$. Models are shown for a \texttt{VapoRock} input surface temperature of 3000~K (orange lines) and 2513~K (dark blue lines, i.e., the sub-stellar temperature of HD~3167~b, as used in Figure \ref{fig:HD3167b_results}). The resulting differences in the relative abundances of the rock vapor species make little difference to the resulting temperature profile and spectrum, especially compared to the more significant effects of adding volatiles to the atmospheric composition.}
    \label{fig:Tsurf_comparison}    
\end{figure*}

In our atmospheric model, \texttt{VapoRock} is used to determine the relative abundances of rock vapor species in the atmosphere, as described in Section \ref{sec:opacity}. The input surface temperature used by \texttt{VapoRock} can be set independently from the equilibrium temperature profile calculated by the radiative-convective atmospheric model. We find that differences in the \texttt{VapoRock} input surface temperature have little effect on the resulting temperature profile and emission spectrum of the atmosphere. This is shown in Figure \ref{fig:Tsurf_comparison} for the example of HD~3167~b. The orange and dark blue lines show temperature profiles and emission spectra for a \texttt{VapoRock} input surface temperature of 3000~K and 2513~K, respectively. The differences between them are negligible, especially compared to the effects of adding volatiles to the atmoshperic composition, which is the focus of this work. For simplicity, we therefore set the \texttt{VapoRock} input surface temperature equal to the sub-stellar temperature of the planet (i.e., 2513~K in the case of HD~3167~b).

\bibliography{refs}{}
\bibliographystyle{aasjournal}



\end{document}